# Applications of brain imaging methods in driving behaviour research

Milad Haghani[1], Michiel C. J. Bliemer[1], Bilal Farooq[2], Inhi Kim[3], Zhibin Li[4], Cheol Oh[5], Zahra Shahhoseini[6], Hamish MacDougall[7]

1. *Institute of Transport and Logistics Studies, The University of Sydney Business School, The University of Sydney, NSW, Australia*
2. *Laboratory of Innovations in Transportation, Ryerson University, Toronto, Canada*
3. *Institute of Transport Studies, Department of Civil Engineering, Monash University, VIC, Australia*
4. *School of Transportation, Southeast University, Nanjing, China*
5. *Department of Transportation and Logistics Engineering, Hanyang University, Republic of Korea*
6. *Department of Transport, VIC, Australia*
7. *School of Psychology, Faculty of Science, The University of Sydney, Sydney, Australia.*

**Abstract**

Applications of neuroimaging methods have substantially contributed to the scientific understanding of human factors during driving by providing a deeper insight into the neuro-cognitive aspects of driver brain. This has been achieved by conducting simulated (and occasionally, field) driving experiments while collecting driver brain signals of certain types. Here, this sector of studies is comprehensively reviewed at both macro and micro scales. Different themes of neuroimaging driving behaviour research are identified and the findings within each theme are synthesised. The surveyed literature has reported on applications of four major brain imaging methods. These include Functional Magnetic Resonance Imaging (fMRI), Electroencephalography (EEG), Functional Near-Infrared Spectroscopy (fNIRS) and Magnetoencephalography (MEG), with the first two being the most common methods in this domain. While collecting driver fMRI signal has been particularly instrumental in studying neural correlates of intoxicated driving (e.g. alcohol or cannabis) or distracted driving, the EEG method has been predominantly utilised in relation to the efforts aiming at development of automatic fatigue/drowsiness detection systems, a topic to which the literature on neuro-ergonomics of driving particularly has shown a spike of interest within the last few years. The survey also reveals that topics such as driver brain activity in semi-automated settings or the brain activity of drivers with brain injuries or chronic neurological conditions have by contrast been investigated to a very limited extent. Further, potential topics in relation to driving behaviour are identified that could benefit from the adoption of neuroimaging methods in future studies.

**Keywords:** driver brain activity; simulated driving; alcohol and cannabis; driving under secondary task; driver decision-making

## 1. Introduction

Road accidents are a major cause for loss of life worldwide and are responsible for more years of life lost than most of human diseases (Petridou and Moustaki 2000). According to the World Health Organisation (WHO), nearly 1.35 million people lose their lives every year as a result of road traffic accidents, while more than half of this figure being related to the death of vulnerable road users. Automobile traffic crashes are also the leading cause of mortality for children and young adults aged 5-29 years. Road trauma costs most countries an equivalent of about 2-3% of their Gross Domestic Product (GDP) in financial terms. For example, according to a study commissioned by the Australian Automobile Association (AAA) (2017), in 2015 and 2016, road accidents were estimated to have cost Australian economy 30 and 33 billion dollars, respectively, which is approximately the equivalent of 2% of the Australian GDP and a figure comparable to its average annual GDP growth rate. These values are based on an estimate of $4.1 million as the Statistical Value of Life (de Blaeij *et al.* 2003, Hensher *et al.* 2009), and reflect various components of the societal cost associated with road accidents – born directly by the individual(s) involved in the accident and society as a whole – including the costs related to life and wellbeing, vehicle damage, disability care, hospital and medical care, travel delay, insurance administration, and emergency services, to name few.

Given the major societal impact of road accidents, an abundance of research has been conducted to better understand the causes of road crashes and effective ways to mitigate them. It is believed that in three out of five traffic crashes, *human factors* and *human errors* are a dominant cause (Adanu *et al.* 2017) and that driver-related factors contribute to the occurrence of 95% of all accidents (Petridou and Moustaki 2000). As a result, a significant portion of the scientific efforts has focused on the element of human factors in driving. This includes a broad array of issues that impact on driver behaviour and ability to perform the tasks of driving including the use of recreational substances such as alcohol (Peck *et al.* 2008, Lenné *et al.* 2010, Mann *et al.* 2010, Irwin *et al.* 2017, Vollrath and Fischer 2017, Yadav and Velaga 2019) and cannabis (Lenné *et al.* 2010, Downey *et al.* 2013, Bondallaz *et al.* 2016), driver's engagement in secondary tasks such as mind wandering (Yanko and Spalek 2014, Geden and Feng 2015), conversing on the phone (Caird *et al.* 2008, Oviedo-Trespalacios *et al.* 2016, Choudhary and Velaga 2017, Lipovac *et al.* 2017) or with passengers (Charlton 2009). In addition to the role of external factors, studies on human factors have also explored the role of individual factors including driver's sociodemographic characteristics (Zhang *et al.* 2019), personality traits (Deffenbacher *et al.* 2002, Bogdan *et al.* 2016, Deffenbacher *et al.* 2016, Demir *et al.* 2016, Sârbescu 2016, Tao *et al.* 2017), style of driving (Taubman-Ben-Ari *et al.* 2004), and driver's mental health, in particular, the effect of high-prevalence psychiatric conditions such as attention-deficit/hyperactivity disorder (ADHD) (Reimer *et al.* 2010, Vaa 2014) or depression (Aduen *et al.* 2015, Aduen *et al.* 2018).

It is well recognised that driving is a complex activity entailing a range of physical as well as mental cognitive tasks. Driving engages various cognitive abilities including decision-making, selected and divided attention, memory, problem solving and planning. It also requires coordination between visuo-spatial attention, visuo-motor skills, perceptual motor skills and auditory skills (Palmiero *et al.* 2019, Ware *et al.* 2020). It is believed that more than forty different tasks are associated with driving among which the response to hazards, lane keeping, speed control and maintaining a safe distance are the primary tasks (Navarro *et al.* 2018). Various conceptual and psychological models of driver decision-making have been proposed, among which the three-level hierarchical model of Michon (1985) – i.e. strategic (planning), tactical (manoeuvring), and operational (control) – appears to be the most prevalent model to date.

The growing application of the neuro-cognitive methods in road safety research is to some extent analogous to the emergence of the field of neuroeconomics (Yu and Zhou 2007), as an integration between neuroscience and economic decision sciences. This integrative approach between behavioural economics and neuroscience as well as the recent scientific developments in that area have helped uncover the underlying mental and neural processes of economic decision making (Sanfey *et al.* 2003, Heekeren *et al.* 2004, Braeutigam 2005, Kenning and Plassmann 2005, Sanfey *et al.* 2006, Clithero *et al.* 2008, Rustichini 2009, Witt and Binder 2013).

Similar to decision sciences, the emergence of advanced neuro-cognitive methods has offered new possibilities for advancing our understanding of human factors in the context of driving behaviour as well. This has pushed the literature of road safety research towards an ever more interdisciplinary nature and has allowed research in this domain to go deeper than the level of merely describing driving behaviour based on external manifestations. It has essentially allowed researchers to better investigate the cognitive mechanisms of driving behaviour through measures of neural activity in a driver's brain. These applications have created a new and rapidly developing sector of research in studies of driving behaviour and have opened new and unique avenues of research that can improve road safety. These methods have shown the potential to predict driver's mental state and fatigue (Barua *et al.* 2019), or to predict their tactical and operational decision-making and intentions (Hernández *et al.* 2018) in a variety of contexts such as emergency braking (Haufe *et al.* 2011, Haufe *et al.* 2014), turning (Zhang *et al.* 2015), acceleration (Vecchiato *et al.* 2019) or driver reactions to Variable Message Signs (VMS) (Yamamoto *et al.* 2019). They have helped better describe the mental mechanisms of mind wandering and its impact on driving performance (Sasai *et al.* 2016), or to discover the neural mechanisms by which alcohol disrupts functional connectivity of driver's brain and impedes performance (Rzepecki-Smith *et al.* 2010), or to better assess fitness to drive for drivers suffering from neurological impairments (Chen *et al.* 2014). This stream of studies on driving behaviour can be regarded as an attempt to open the "black box" of driver's brain (Yu and Zhou 2007, Camerer 2008) and to explore the underlying neural and psychological mechanisms of driving.

Traditionally, the behaviours of drivers have been studied using driving simulators as a safe method that offers great controllability over the design factor(s) of interest and the levels of stimulus as well as an acceptable level of ecological validity (Brooks *et al.* 2010, Navarro *et al.* 2018). Aspirations to gain deeper insights into drivers' brain at the neural and cognitive levels has generated a growing body of studies in which human brain activity is measured while performing (often simulated) driving tasks (Calhoun and Pearlson 2012). As suggested by Graydon *et al.* (2004), while significant progress has been made in identifying underlying behavioural factors associated with driving, still little is known about how the brain executes this complex and multi-task activity. They argued that emerging and constantly upgrading vehicle control systems that continue to increase driver's cognitive attentional demand further highlight the need to further our understanding of how brain systems during driving are modulated, what brain systems are recruited and how these brain systems evolve as a result of driving practice and experience. It is also expected that findings from this stream of neuroimaging driving studies can be applied to better inform the design of efficient and safe vehicle control systems and interfaces.

Here, this particular sector of the road safety and driving behaviour research, i.e. the neuroimaging driving behaviour studies, is reviewed. The aim is to (i) identify which brain imaging methods are being used in this domain and to establish their advantages and shortcomings for various research questions specific to the domain, (ii) identify the most common themes of driving behaviour studies that have made use of brain imaging methods, (iii) synthesise the findings of these studies and contrast them with one another as well as with established findings in the mainstream road safety literature; and (iv) to identify research gaps and underexplored areas where applications of brain imaging methods in driver behaviour experiments have the potential to contribute novel insights. The literature analyses are conducted at a macro scale, where we identify general trends on this topic, as well as at a micro scale, where we analyse findings and methods of individual studies and contrast them.

In the next section, we delineate brain imaging methods applied in road safety research, general themes of research using these methods, as well as our review method and inclusion/exclusion criteria. Section 3 provides a macro-scale analysis of the literature on this topic. In Section 4, individual studies are categorised into clusters of similar themes, individual studies are analysed, and characteristics of the studies are summarised. Section 5 provides summary statistics of the individually analysed studies. Section 6 discusses our findings and provides directions for further research in this domain. Also, some advantages and disadvantages of various brain imaging methods are discussed in the specific context of driving behaviour experiments.

## 2. Methods and data

The overarching theme of this review study is the applications of brain imaging in experiments of driving behaviour, i.e. collecting drivers' brain signals as they perform driving tasks. As will be detailed in Sections 3 and 4, studies of driving behaviour that utilise measures of neural activity of driver's brain have been showing an increasing trend in terms of quantity and have addressed a diverse range of dimensions related to driving behaviour. They are also dispersed across a broad range of scientific sources and multiple disciplines including transportation science, ergonomics, experimental psychology and cognitive neuroscience. Brain signals during driving tasks in these studies have been obtained from four general methods:

(a) *Functional Magnetic Resonance Imaging* (fMRI), or functional MRI, which is a method for depicting changes in deoxyhemoglobin concentration consequent to task-induced or spontaneous modulation of neural metabolism (Glover 2011). Established in 1990, this method has been widely utilised in numerous cognitive, clinical and behavioural studies and since 2001 has been adopted to learn about driver's brain activity. The method was developed to demonstrate regional, time-varying changes in brain metabolism and relies on Blood Oxygen Level Dependent (BOLD) signal. It is based on the premise that cerebral blood flow and neuronal activation are coupled: when an area of the brain is in use, blood flow to that region also increases (Logothetis *et al.* 2001). The method requires that subjects be placed motion-less in an MRI machine as they perform a given task.

(b) *Electroencephalography* (EEG), which is a method to record electrical activity in the brain by measuring voltage fluctuations of the ionic current within neurons of the brain (Cultice 2007). This electrical activity is recorded over a period of time by multiple electrodes placed on the scalp. The method predates fMRI by a long time and has been in use since the 1930's. Applications of this method in driving behaviour research were reported as early as 1978 (Bente *et al.* 1978).

(c) *Functional near-infrared spectroscopy* (fNIRS) is a method that basically uses NIRS for functional neuroimaging and captures the changes in optical properties of brain tissue (Villringer and Chance 1997). Using this method cerebral hemodynamic responses are measured by near-infrared light propagating through the head and gathering information about volume, oxygenation and flow of blood. A sensor is attached to the subject's forehead and connects directly to a computer. It can also connect to a portable computing device that records the signals as the subject performs given tasks (Ferrari and Quaresima 2012).

(d) *Magnetoencephalography* (MEG) is another functional neuroimaging method that records small magnetic fields produced in the brain. Like fMRI, the method requires a scanning machine, but unlike fMRI, an MEG scanner does not emit radiation or magnetic fields (Boto *et al.* 2018).

This work focuses on studies that have used any of these four methods to measure the brain activity of experimental subjects while performing (mostly simulated, and occasionally on-road) driving tasks. The use of Positron Emission Tomography (PET), as another method of measuring brain activity, has also been reported in very few studies (Horikawa *et al.* 2005, Jeong *et al.* 2006), but, since the method has not been prevalent in this domain and since some of PET applications were focused on clinical effects of drugs on cognitive performance in general rather than driving per se (Tashiro *et al.* 2008, Tashiro *et al.* 2009), this review only focuses on the four main methods mentioned above. In the context of driving behaviour, studies that have employed brain imaging methods have predominately used either EEG or fMRI. Much fewer applications of fNIRS and particularly MEG have reported, by comparison. These methods each offer a range of opportunities and challenges for studying driving behaviour that need to be considered and traded off in relation to the research question at hand.

The surveyed literature was systematically searched and retrieved from a variety of data sources, including Web of Science (WoS), Scopus and Google Scholar, though the primary source was WoS. To retrieve a general dataset of references, various combinations of key terms were used in the topic search of the WoS platform (which covers title, abstract, author keywords and extended keywords of the documents) and outcomes of the search were examined. The following combination of the key terms and logical operators (AND, OR) appeared to have returned the most accurate results in terms of sensitivity and specificity:

*("driving* behaviour*"* OR *"simulated driving"* OR *"driving simulation*"* OR *"car driving"* OR *"automated* driving"* OR *"semi-automated* driving"* OR *"driving* distraction"* OR *"driving* fatigue"* OR *"driving* drowsiness*"* OR *"intoxicated driving*"* OR *"traffic psychology")*
AND
*("brain activity"* OR *"brain imaging"* OR *"neural activity"* OR *"neural correlate*"* OR *"functional MRI"* OR *"fMRI"* OR *"EEG"* OR *"Electroencephalography")*

The inclusion of the asterisk sign allows the variations of the words to be detected and included as well. The terms fNIRS and MEG were not included in the search for two reasons. Firstly, the relevant studies in these categories were captured through the rest of the key terms, and secondly, the inclusion of these terms resulted in a high number of false detections, such as documents related to clinical studies. Only a few research outcomes prior to 2000 were identified, and a continuous stream of studies in this domain was published after this year. The search was last time updated in February 2020 where it returned 357 relevant documents. We refer to this as the *general dataset* of references in this work. The full record of these documents, including details of the authors, titles, sources, and the cited references, was exported from WoS and were used for a macro-scale analysis of the literature. The items in this general dataset were subsequently examined one by one and were filtered out based on a set of inclusion and exclusion criteria in order to form a *core dataset* of references for the subsequent micro-scale analysis. Firstly, the documents were filtered based on document type, and only journals articles published by peer reviewed journals were included in the core dataset. Proceeding papers, book chapters, reviews, editorials and meeting abstracts were excluded. The abstract and often full content of the remaining articles were examined to further filter the items. Only studies with major behavioural components related to driving behaviour were considered. The use of brain imaging methods was a necessary but not sufficient condition. Rather, studies need to have had offered behavioural insights via their brain imaging data. Studies with merely a focus on clinical aspects, or technological aspects of the brain mapping or those in which the brain signal was not particularly recorded while driving were excluded. Studies on drivers using non-functional MRI, such as those studying the brain structure of taxi drivers or car racers (Maguire *et al.* 1997, Maguire *et al.* 2000, Maguire *et al.* 2006a, Maguire *et al.* 2006b, Bernardi *et al.* 2013, Lappi 2015) were deemed out of the scope. The search for the core dataset was repeated in the same manner in Scopus. Once a preliminary core dataset was formed, studies were categorised into seven general themes (see Section 4 for details). Within each of these seven categories the topmost cited and the newest items were singled out, and the reference lists of the most recent as well as the most cited articles (based on Google Scholar records) were further examined to identify possible new items for the core dataset. This procedure resulted in a core dataset of 85 items.

## 3. Macro-scale analysis of the literature

The general dataset of references was analysed at the macro scale in order to discover the trends and patterns of publications in the domain of road safety research using neuroimaging methods. Figure 1(a) demonstrates the temporal distribution of the studies published between 2000 and 2019 that were retrieved from WoS. The data shows that the number of publications in this domain has been steadily on the rise since 2000, but the rate of publication has become steeper since 2013. The distribution of these studies across WoS categories as well as their distribution across various journals and conferences (exclusive to those with at least 5 items in this domain) have respectively been visualised in Figures 1(b) and 1(c). As indicated by these figures, the related studies have been distributed across a large variety of fields including neuroscience, engineering, psychology, transportation and ergonomics. In terms of the number of publications, studies published by *Accident Analysis and Prevention* and *Frontiers in Human Neuroscience* have contributed most to this domain.

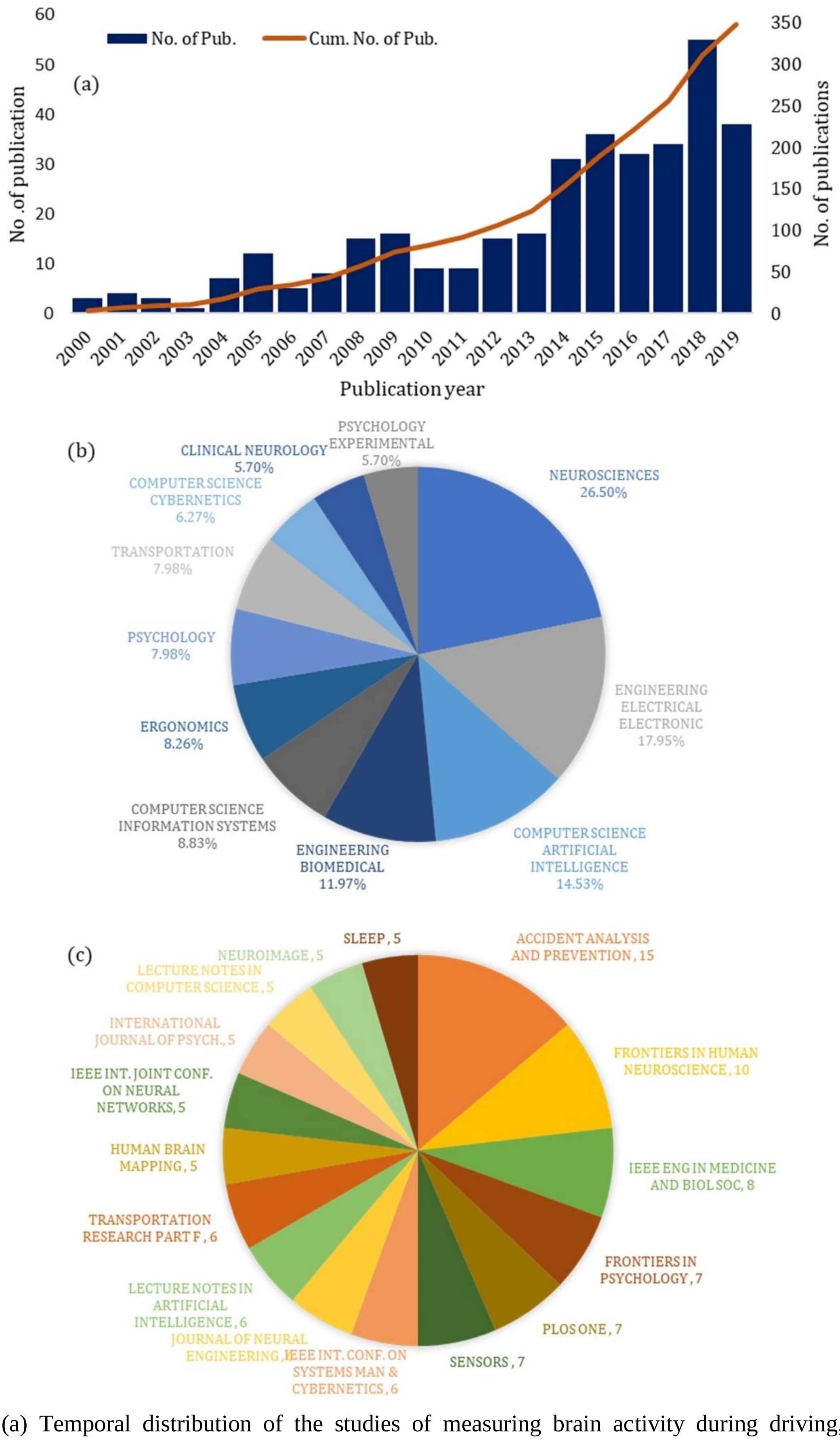

**Figure 1** (a) Temporal distribution of the studies of measuring brain activity during driving, (b) their distribution across fields of studies and (c) their distribution across major journals and conferences. In (c), numbers represent the number of published items of each source on this topic.

The citation patterns of the publications on neuroimaging studies of driving behaviour were also analysed based on the dataset of references obtained from WoS. Figures 2(a) and 2(b) respectively provide the map of co-citation and the map of bibliographic coupling of the sources in the general dataset. The map has been constructed using open access software VOSviewer (Van Eck and Waltman 2013), which is specialty software for scientometric analysis.

Two publications are regarded co-cited in the dataset when they have been cited together by a same third publication, and the associated map of co-citations in Figure 2(a) represents this relationship between the publications aggregated over the source (e.g. journal) of origin. In order to maintain clarity of the map, the minimum number of citations for a source to appear on this map has been set to 20. The size of each circle (or node) associated with each source in the map is proportional to the number of citations received by the studies of that source on our topic of interest. The thickness of the links between the nodes represents the strength of the co-citation relationship. Sources with strong co-citation relationships generally appear spatially closer together, therefore clusters of co-cited journals can be identified, and we associated different colours with different clusters. In this case, three distinct clusters of co-cited journals can be identified. The blue cluster is mainly represented by *Accident Analysis and Prevention* whose publications (despite receiving many citations) on this topic appear to be only highly co-cited with *Sleep*, *Journal of Sleep Research*, *Human Factors*, *Traffic Injury Prevention* and *Physiology & Behaviour*. The green cluster of co-cited journals on our topic of interest mainly consists of journals in neuroscience and cognitive neuroscience. The most distinct source within this cluster is *Neuroimage* which appears to have relatively strong co-citation relationships with *Human Brain Mapping* and *Transportation Research Part F* (from within that cluster) as well as with *Accident Analysis and Prevention* from the blue cluster. The red cluster shows the least within-cluster strength of co-citation relations among the three main clusters.

Bibliographic coupling is another measure of the relatedness of publications and represents the number of references that two sources share (Marshakova 1973). The higher the number of shared references between two studies, the stronger their bibliographic coupling. This relationship is visualised in Figure 2(b). The minimum number of publications for sources to appear in this map has been set to 3. Here, sources with a larger number of publications on our topic of interest appear more distinctly. The thickness of links connecting their corresponding nodes is proportional to the strength of their bibliographic coupling, i.e. the number of references that their publications on this topic have in common. Here also three main clusters of bibliographically coupled sources were identified with *Accident Analysis and Prevention* being the most prominent source according to its high number of published items on this topic. Its publications on this topic show a strong bibliographic coupling with those of *Biological Psychology*, *Sleep*, *Physiology & Behaviour* from within the same cluster of coupled sources as well as with sources such as *Frontiers in Human Neuroscience* and *Frontiers in Psychology* from other clusters.

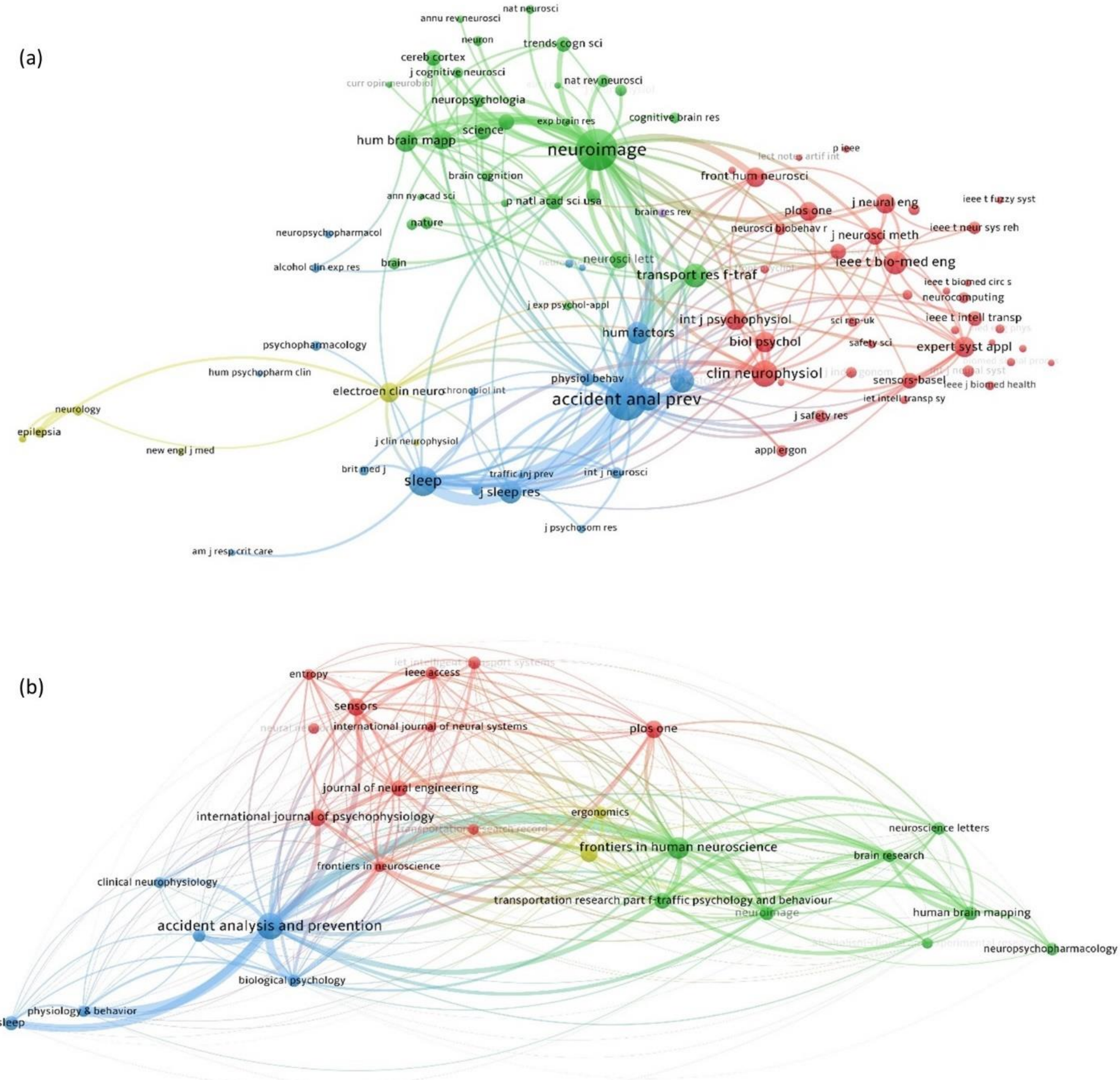

**Figure 2** The maps of (a) co-citation and (b) bibliographic coupling of the sources based on their publications on neuroimaging applications in driving behaviour research.

To generate a better macro-scale picture of the literature on neuroimaging studies of driving behaviour, texts of the title, abstract and keywords of the publications on this topic were also analysed using the same software. Figure 3(a) shows a map of co-occurrence based on the analysis of keywords, and Figure 3(b) shows a map of term co-occurrence based on the analysis of titles and abstracts. The nodes in each map correspond with a specific term and their sizes represent the frequency of occurrence of the corresponding term. A link between two nodes indicates the co-occurrence of their corresponding terms in publications. Terms that co-occur frequently appear relatively closer to each other on the map and can form clusters. These clusters have been represented by different colours in Figure 3(a). In Figure 3(b), however, the clusters have been forgone and instead, the map has been overlaid with a colour-coding that represents the average of the publication years associated with the studies in which each term has appeared. Figure 4 provides the corresponding maps to those of Figure 3, in a heatmap format, where darker colours (i.e. closer to red) correspond with higher occurrences of the terms. It shows various clusters of the high-occurring terms based on keywords search (Figure 4(a)) or title and abstract text search (Figure 4(b)) without the interference of the links and nodes.

At first glance, the maps (in Figures 3 and 4) reflect prominent generic terms such as "*driving performance*", "*simulated driving*", "*simulated car*", "*laboratory", "road*", "*drivers*", "*car driver*", "*traffic safety*" as well as terms associated with measuring driving performance such as "*lane deviation*", "*deterioration*", "*stimuli*". According to both analyses, the term "*eeg*" and its associates like "*eeg signal*" have been the most common terms among these studies. This is a reflection of the EEG method being the most commonly used method of measuring brain signals in driving behaviour experiments.

In Figure 3(a), the red cluster on the left end of the map represents keywords that are affiliated with "*fMRI*" applications in driving behaviour studies. This includes terms such as "*brain activity/activation*", "*brain function*", "*human brain*", "*brain network*", "*event-related potentials*", "*neural*", "*(functional) connectivity*", and "*neuroergonomics*" that are generally affiliated with fMRI experiments. Terms such as "*visual*" and "*motor*" represent to some degree the variety of functionalities of driver's brain analysed in these studies. This cluster also includes terms referring to various localised brain areas that have found to be relevant in executing driving tasks. This includes terms such as "*cortex*", "*prefrontal cortex*", "*parietal*", "*anterior cingulate cortex*", "*basal ganglia*", "*cerebellum*", "*anterior cingulate cortex*" and "*dorsolateral prefrontal cortex*" (see Hawrylycz *et al.* (2012) for an anatomically comprehensive atlas of the adult human brain). Furthermore, terms associated with fMRI data analysis and fMRI experiment designs are noticeable within this cluster. This includes terms such as "*event-related potentials*", "*fMRI time series*" and the particularly distinct term "*independent component analysis*", representing the method that was adopted, as an alternative to the more common method of General Linear Model, by Vince Calhoun and his team (Calhoun and Pearlson 2012) for the analysis of the fMRI signal in experiments of continuous driving with non-event-related designs where the onset and offset of driver actions cannot be predicted a priori. The term "*virtual reality*" also appears in this cluster that represents a pragmatic technological alternative to more sophisticated versions of simulated driving that researchers devised in order to overcome technical challenges of driving experiments in fMRI settings and to make the integration between driving and fMRI brain imaging a possibility in an ecologically valid fashion (Calhoun *et al.* 2005). Terms such as "*ethanol*", "*alcohol*", "*blood alcohol concentration level*" (BAC), "*placebo*" and "*performance impairment*" in this cluster are reflections of a cohort of studies published largely between 2004-2010 focusing on the neural mechanisms of alcohol affecting driver brain networks (Calhoun *et al.* 2004a, Calhoun *et al.* 2005, Allen *et al.* 2009). The term "*spatial attention*" also appears in this cluster, which could be related to another group of studies that adopted the fMRI method for experiments focused on investigating a distracted brain when driving (Graydon *et al.* 2004, Just *et al.* 2008). On the fMRI end (i.e. the left-hand side) of the map in Figure 3(b) (or its corresponding heatmap in Figure 4(b)), the term "*decision making*" appears in close proximity to fMRI. This is a reflection of a group of studies that have adopted the fMRI method to study the neural correlates associated with various aspects of driver decision-making in uncertain scenarios and, in particular, risk-taking behaviour (Callan *et al.* 2009, Chein *et al.* 2011). According to the same map, the term "*healthy volunteer*" appears close to the terms associated with fMRI which essentially is a reflection of the fact that the vast majority of fMRI driving behaviour experiments have focused on the behaviour of healthy (and often young) adults while fMRI studies that have focused on the brain activity of older drivers or those with brain injury or chronic conditions are very rare comparatively (Hung *et al.* 2014).

Adjacent to the red cluster in Figure 3(a) is a blue cluster whose terms are generally attributable to the neuroimaging studies of driving while performing a secondary task, i.e. distracted driving (Graydon *et al.* 2004). The cluster includes terms such as "*attention*", "*visual-attention*", "*telephone*", "*interference*", "*driver distraction*", "*dual-task*" and "*cognitive control*", which are some generic terms that characterise this cohort of studies. It is worth noting, however, that within this cluster, a mixture of terms affiliated with both fMRI and EEG measurements (e.g. "*eeg control*", "*eeg alpha*", "*prefrontal cortex*", "*neuroimaging*") have appeared which demonstrates the fact that both methods have been adopted by studies on distracted driving (Bowyer *et al.* 2009, Hsieh *et al.* 2009, Uchiyama *et al.* 2012, Schweizer *et al.* 2013, Choi *et al.* 2017).

Moving towards the right-hand side of the map in Figure 3(a), terms become more and more affiliated with the topic of driver fatigue and sleepiness. This includes the yellow and green clusters in the middle as well as the light blue cluster at the other end of the map. The yellow cluster is represented by terms such as "*sleep*", "*driver drowsiness*", "*driver sleepiness*", "*deprivation*", "*obstructive sleep apnea*", "*alertness*", "*lapses*", "*vigilance*", "*falling asleep*", "*caffeine*", "*awareness*", which are all clearly representative of studies of driving in a drowsy state. Next to this cluster, terms associated with the studies of "*(driver) fatigue*" have appeared that also seem to have strongly co-occurred with the term "*eeg*". This is mainly due to the fact that studies of this category have exclusively been conducted using the EEG method for measuring brain activity (Kar *et al.* 2010, Huang *et al.* 2015, Perrier *et al.* 2016). As will be discussed in the following sections in more detail, these studies have predominantly focused on the development of "*brain-computer interface*" systems that can be used for the assessment and "(*early*) *detection*" of "*mental/driving fatigue*" or "*braking*" intentions using "*eeg*" signal, e.g. "*driver drowsiness detection*" systems. These studies have used various "*machine learning*" and "*deep learning*" methods to identify patterns of electrical "*brain activity*" that match other indicators of driver "*sleepiness/fatigue*" such as "*heart-rate (variability)*", and "*eye (movements)*"/closure or "*respiration*". Technical terms such as "*classification*", "*convolutional neural networks*", "*support vector machine*", "*entropy*", "*fuzzy entropy*", "*spectral entropy*", "*fuzzy entropy*", "*wavelet entropy*", "*approximate entropy*", "*sample entropy*", "*power spectrum analysis*" associated with the analysis of EEG signal are reflections of this cohort of studies on the neuro-ergonomics of driving (Gurudath and Riley 2014, Chuang *et al.* 2015).

An inspection of the map in Figure 3(b) also shows that the term "*healthy subject*" has also co-occurred with the terms associated with EEG. In conjunction and consistent with the co-occurrence of the term "*healthy volunteers*" with fMRI-associated terms (as mentioned earlier), these patterns collectively demonstrate how these studies have predominantly been dedicated to the investigation of a healthy drivers' brain with much less attention to the neuro-cognitive characteristics of drivers post brain trauma or those with chronic neurological conditions.

Given that EEG as a general method of measuring brain activity predates fMRI, one might assume that studies of neuroimaging of driver brain using fMRI must be more recent compared to the counterpart EEG studies. But as Figure 3(b) clearly shows, the reverse is the case. Both this macro-scale analysis and the subsequent micro-scale analyses that will be presented in the following sections confirm that in recent years, the body of studies on neuro-ergonomics of driving has predominantly focused on EEG applications particularly in automatic driver fatigue/sleepiness detection systems. While the appearance terms affiliated with fMRI studies are centred around the years preceding 2012, the appearance of the EEG terms have mostly occurred from 2014 onwards with terms "*sample/approximate/spectral entropy*" or "*clustering coefficient*" being the youngest in this domain.

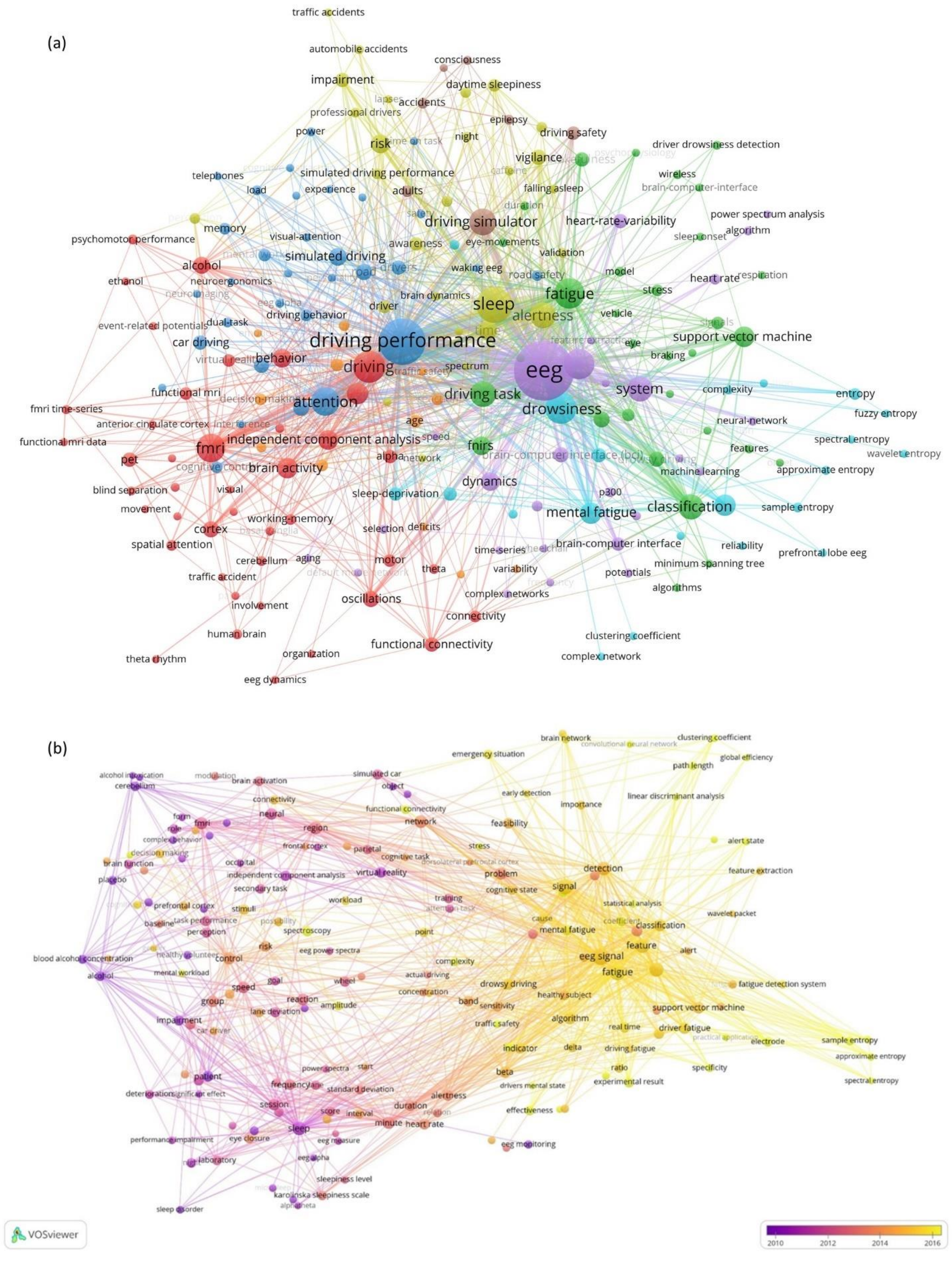

**Figure 3** The map of co-occurrence of (a) keywords and (b) the map of co-occurrence of the terms in the title and abstract of publications in studies of neuroimaging applications in driving behaviour research. In (b), the map has been overlaid with a colour-coding of the average year of associated with each term.

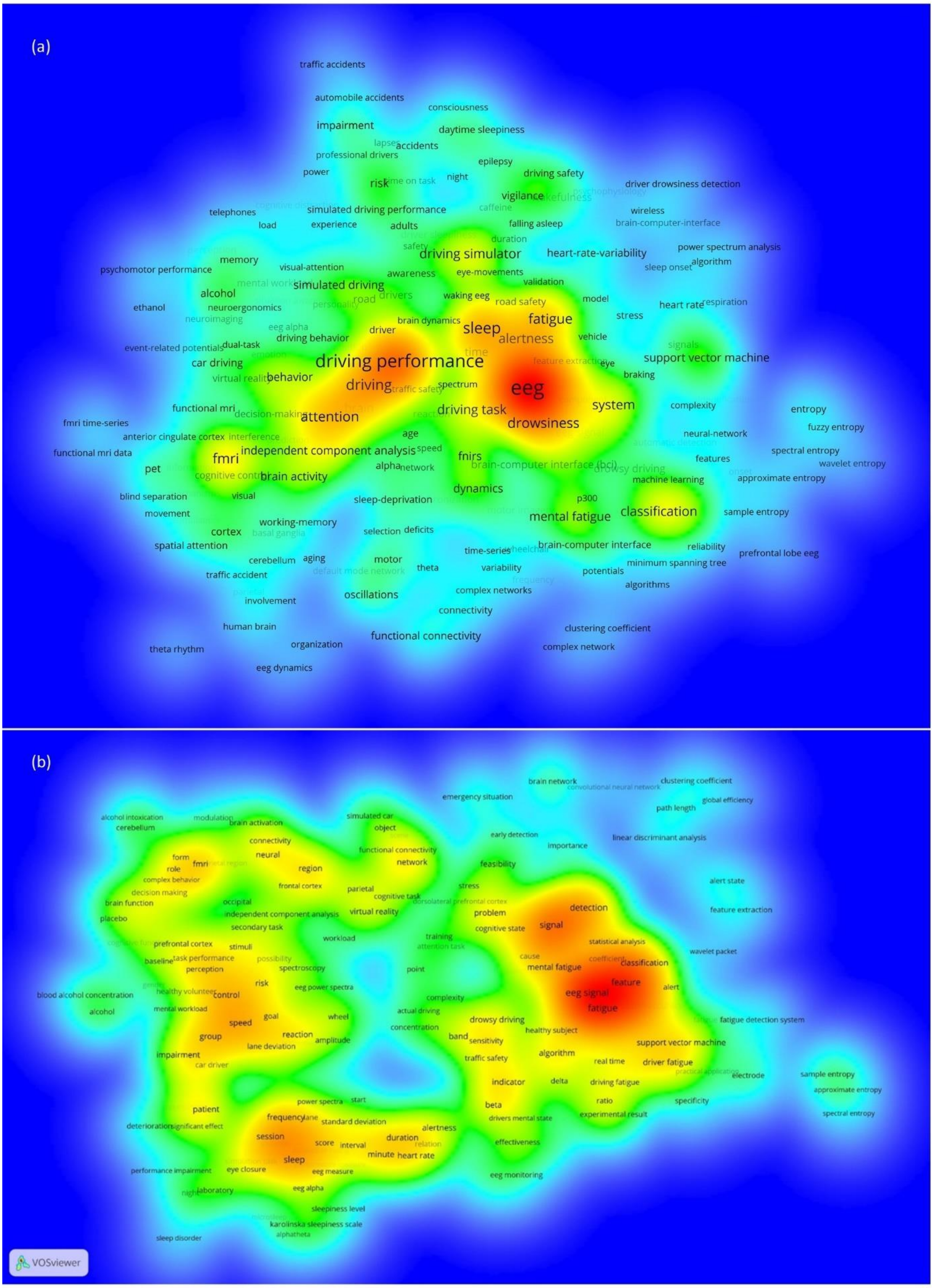

**Figure 4** The map of co-occurrence of (a) keywords and (b) the map of co-occurrence of the terms in the title and abstract of publications in studies of neuroimaging applications in driving behaviour research, in the form of heatmaps.

## 4. Micro-scale analysis of the literature

After analysing and categorising individual studies of driver brain imaging, seven major themes of research related to driving behaviour were identified to have been investigated using brain imaging methods. These include (indicating the number of studies in square brackets):

1. Behaviour of drivers under the effect of substances (alcohol, cannabis) [9];
2. Behaviour of distracted drivers (conducting secondary tasks) [14];
3. Behaviour of drivers with brain impairment and underlying neurological problems [5];
4. Behaviour of drivers in semi-automated settings [4];
5. Behaviour of drivers under the effect of fatigue or sleep deprivation [21];
6. Drivers' decision-making and risk-taking behaviour [11];
7. General driving behaviour of healthy subjects in a sober state without distractions or automation [21].

In the following sub-sections, studies of each category are discussed separately while their details including main research questions, study design, brain imaging findings and behavioural findings have been summarised in Appendices A to G.

### *1. Brain activity of intoxicated drivers*

Driving while inebriated remains a major cause of traffic accidents and has conventionally been a main focus of attention in road safety research. Driving simulator experiments have almost invariably been the primary tool for studying intoxicated driver behaviour, while using various latitudinal and longitudinal measures of driving behaviour as metrics of driver performance while under the influence of alcohol (Irwin *et al.* 2017). A similar approach has also been taken in relation to studies of drivers under the influence of recreational drugs such as cannabis. Validation studies have produced evidence of ecological validity of the driving simulator to study the behavioural effect of recreational drugs (McGinty *et al.* 2001).

Following the study of Walter *et al.* (2001) on general neural correlates of driving with healthy drivers, which first adopted the fMRI method in driving experiments domain, Vince Calhoun and his team pioneered a series of studies on brain functional connectivity of healthy drivers (Calhoun *et al.* 2002) as well as that of intoxicated drivers (i.e. under acute alcohol administration) (Calhoun *et al.* 2004a) using fMRI signal as an extra layer of measurement in simulated driving experiments. This stream of studies aimed at understanding how neural substrates of driving are affected by alcohol intoxication (Calhoun *et al.* 2004a, Calhoun *et al.* 2005, Calhoun and Pearlson 2012). They implemented simplified versions of virtual-reality simulated driving which allowed subjects to perform the tasks while positioned in an MRI machine. Their experiment designs often involve blocks of fixation on target, followed by simulated active driving and followed by watching or passive driving.

Cognitive brain imaging studies often rely on the subtraction of the BOLD signal between two types of tasks in slight increments. Having subtracted the BOLD signal, they usually visualise the brain regions that differ. In their studies, they highlight the difficulties of analysing temporal dynamics of driving with fMRI, considering the lack of a well-understood brain activation model. They argued that during driving, as a complex cognitive task, multiple brain circuits could be activated simultaneously where a particular region may contribute differentially to multiple circuits. Also, multiple driver response could overlap in time. They suggested that these concerns all make the use of conventional fMRI data analysis methods rather questionable for studying driving tasks, where the onset and offset of driver actions cannot be known a priori (Calhoun and Pearlson 2012). In order to be able to study temporal event-related dynamics of driving, Calhoun and colleagues pioneered the use of the Independent Component Analysis (ICA) method (McKeown *et al.* 2003, Daubechies *et al.* 2009) in their studies of sober and intoxicated driving behaviour. This method was favoured as opposed to the more prevalent General Linear Model (GLM) (Calhoun *et al.* 2004b) method considering that GLM cannot detect brain activity with time courses not known in advance. This exempts the analyst from having an a priori hemodynamic model and enables them to study temporal dynamics (Erhardt *et al.* 2011). In their approach, they "assume independence of the hemodynamic source locations from the fMRI data (independence in space) resulting in maps for each of these regions, as well as the time course representing the fMRI hemodynamics" (Calhoun *et al.* 2002) (p. 159). Using the ICA approach, they produced subject-specific maps and time courses that can be contrasted to time courses of various behavioural measures recorded during the simulated driving task. A hybrid of ICA and GLM has also been reported in their studies, though in a distracted while intoxicated driving application, where additional tasks (such as acknowledging a salient stimulus within the virtual vehicle) is timed a priori (Allen *et al.* 2009). That study uses a dual-task paradigm in the experiment design which involves performance of a visual oddball (VO) task (Stevens *et al.* 2000) while driving in an alcohol challenge paradigm.

In terms of inter-network brain connectivity, these studies have collectively identified five independent brain circuits whose connectivity is disrupted by acute alcohol consumption. These studies have also found significant signal changes in the orbitofrontal (OF) anterior cingulate cortex of the brain when the driver is under the effect of alcohol (Calhoun *et al.* 2004a). Also, dose dependent signal changes have been revealed in OF and motor regions as well as visual and medial frontal regions. In light of these observations, it has been suggested that the attentional deficits may be mainly modulated by the OF and cerebral regions rather than the attentional areas in the frontoparietal (FP) cortex. These studies have indicated impairment of the error-detection brain regions as a result of the alcohol intoxication (Calhoun *et al.* 2005)

In terms of behavioural measures, and consistent with the main body of conventional driving simulation studies, these studies have demonstrated the association between alcohol dose and increased collision and near collision (at a higher dose), less responsiveness to peripheral events, depressed perceptual and motor functioning (Calhoun *et al.* 2005), increased reaction time and lesser ability to execute secondary tasks (Allen *et al.* 2009), increased steering weave (particularly when combined with sleep deprivation (Vakulin *et al.* 2007)) and increased frequency of line crossing events on the passenger side (Meda *et al.* 2009). In contrast, some studies have found that a low alcohol dose is associated with more careful driving, lower speeds and slightly improved performance (Calhoun *et al.* 2004a, Allen *et al.* 2009, Carvalho *et al.* 2014).

A smaller number of studies in this domain investigated the neural effects of cannabis on driving as another form of a recreational intoxicant (Battistella *et al.* 2013, Brown *et al.* 2019). The study of Battistella *et al.* (2013) used an fMRI paradigm and visuo-motor tracking task. The design entailed active tracking blocks, passive tracking viewing blocks and rest. Their observations demonstrated reduced BOLD signal in the anterior insula, dorsomedial thalamus, striatum, right superior parietal cortex and dorsolateral prefrontal cortex of drivers under the effect of cannabis compared to the placebo condition. In addition, cannabis consumption resulted in increased activity in brain regions associated with self-oriented mental activity. The study of Brown *et al.* (2019) used EEG to investigate acute cannabis intoxication and its effect on brain activity. They observed a significant correlation between impaired latitudinal driving performance and EEG power in slow theta band in the parietal and occipital areas. Their investigation revealed several biomarker candidates associated with cannabis ingestion that can be derived from EEG signals. Details of the studies on the brain activity of intoxicated drivers have been synthesised in Appendix A.

### 2. *Brain activity of distracted drivers*

Driving while performing a secondary task, i.e. dual/multiple-task driving, has been a major focus of road safety research (Nasar *et al.* 2008, Zhang *et al.* 2019, Karthaus *et al.* 2020, Ma *et al.* 2020). Studies on the neural correlates of drivers with a distracted mind are also well represented in the literature surveyed in this review. These studies have investigated the effect of a range of distracting/secondary stimuli on driver brain activity, including the effect of sentence comprehensions and listening to conversation/speech (Just *et al.* 2008, Fort *et al.* 2010, Uchiyama *et al.* 2012, Schweizer *et al.* 2013), covert conversation (Bowyer *et al.* 2009, Hsieh *et al.* 2009), visual event detection or visual vigilance (Graydon *et al.* 2004, Al-Hashimi *et al.* 2015, Xu *et al.* 2017), simple arithmetic calculations (Chung *et al.* 2014, Choi *et al.* 2017), and mind wandering (Lin *et al.* 2016, Baldwin *et al.* 2017).

Previous research has extensively investigated the performance of drivers while engaged in secondary auditory tasks such as hands-free phone conversations, speaking with in-car passengers and speaking with remote passengers (Charlton 2009). Neuroimaging studies have provided more insight into the effect of auditory language comprehension on neural mechanisms of driver brain activity. Just *et al.* (2008) argue that while multitasking and driving and conversing on a mobile phone has been made technologically feasible, its impact on driving performance is still debated and requires more investigation to establish what aspects of driving are likely to be affected and to what extent. They conducted experiments of auditory language comprehension while simultaneously performing a simplified simulated drive in an MRI scanner. The auditory task required participants to determine whether common knowledge sentences presented to them were true or false and the design included blocks of driving alone followed by driving while listening. They observed that with the addition of the sentence comprehension task, brain activation in the parietal and superior extrastriate decreased while the activity in the temporal and prefrontal language areas increased. They concluded that while the driving and language comprehension tasks largely draw on non-overlapping cortical areas, the introduction of the secondary task significantly decreases the activity of brain areas associated with spatial processing during driving. This could be an indication of reduced spatial computations as well as spatial attention and shows that engaging in auditory task draws mental resources away from driving, which could explain the observed deterioration in driving performance. In another fMRI study, Schweizer *et al.* (2013) included an auditory task analogous to a hands-free phone conversation (i.e. true or false common knowledge sentences) in a simulated driving experiment. They also observed that the distracted brain sacrifices activity in the posterior visual and spatial areas important for visuo-spatial processing to recruit more resources in the prefrontal cortex to perform the secondary auditory task. Using functional MRI, Uchiyama *et al.* (2012) also investigated changes in brain activity during a car-following task engendered by a concurrent auditory task. The auditory task entailed language comprehension and tone discrimination. They observed a decrement in the car following performance during the dual task while also observing a suppression of brain activity in regions that are important in performing the car-following task. Sasai *et al.* (2016) studied functional brain split (Gazzaniga 2014) in a simulated driving listening paradigm where drivers engaged in either an integrated task (i.e.

listening to the GPS) or a split task (i.e. listening to a radio) as fMRI data was collected. Their experiment demonstrated that a driver's brain may functionally split into two separate driving and listening systems when the listening task is unrelated to the driving task, but not when the two tasks are integrated. The authors made an analogy between this finding and similar observations in the field of neuroscience suggesting that after surgically disconnecting the two cerebral hemispheres in certain patients to reduce epileptic seizures, patients can still continue to function normally, or that after a split brain operation two separate streams of consciousness, one per hemisphere, can coexist in one brain (Gazzaniga 2014).

In two linked studies, the effect of conversations (as the secondary task) was investigated in the neural mechanisms underlying reaction time of drivers using fMRI (Hsieh *et al.* 2009) and MEG (Bowyer *et al.* 2009). Subjects in these experiments performed covert conversations while also tasked with a vidual vigilance activity during passive simulated driving. In both studies, the dual task resulted in longer reaction times associated with visual event detection though the effect on miss rate was negligible. A frontal-parietal network was identified that maintains event detection performance during the conversation task. During the dual task, activity was observed to increase in the Broca's and Wernicke's language regions in addition to a number of other brain areas (Hsieh *et al.* 2009).

In a limited form of simulated driving in an MRI machine, Graydon *et al.* (2004) observed that a visual event detection task during driving engages multiple interconnected cortical and sub-cortical neural systems. They also identified a more prominent role for fronto-parietal networks during simulated driving that require greater attention demands associated with the visual event detection task. In an effort to increase the ecological validity of the simulated driving, their experiments used a real-world driving scene (video) rather than a computer-generated scene. Their visuo-motor task required drivers to respond by pressing a button to a target visual stimulus that appeared in a predictable location but at a random time. In a video-game-like simulated driving experiment and using the fMRI method, Al-Hashimi *et al.* (2015) also assessed the neural correlates of multi-tasking (visual event stimuli) performance decline with subjects in their 30's and 40's. They observed that the activity of only a single brain region, the superior parietal lobe, showed a significant correlation with multitasking performance.

Driving in a distracted state could, of course, manifest itself in the absence of any external stimuli or secondary task, and could present itself in the form of mind wandering (Walker and Trick 2018, Burdett *et al.* 2019). According to Albert *et al.* (2018) "Mind wandering, which encompasses thoughts and feelings unrelated to ongoing tasks, is an internal form of distraction that also has significant traffic safety implications" (p. 126). The study of Baldwin *et al.* (2017) reports on recurring experiments of monotonous simulated freeway driving while collecting EEG brain signals and periodically probing drivers to self-report their state of distraction. They observed that the frequency of mind wandering was high among drivers, its frequency did not significantly increase over the days of participation, but it did increase during the second drive compared to the first drive within the same day. They also observed that periods of mind wandering were associated with increased EEG power in the alpha band. Another EEG-based simulated driving experiment reported by Lin *et al.* (2016) showed that mind wandering during driving tends to occur under low perceptual demand. Details of the studies in this category have been summarised in Appendix B.

### 3. *Brain activity of drivers with brain impairment and underlying neurological illness*

Driving brings many people the benefits of mobility and access and is, therefore, linked to their economic and social wellbeing. As a result, even people with chronic illnesses or disabilities will seek to gain a driving license or to maintain their licenses for as long as possible even after serious injury, illnesses or advancing age (Unsworth and Baker 2014). On the other hand, driver's health is a crucial prerequisite as the association between suffering from chronic medical conditions or injuries and the increased risk of road crashes is well established (Marino *et al.* 2013) . Therefore, from a regulatory perspective, this makes the evaluation of fitness to drive (Baker *et al.* 2015) an important matter for people recovering from injuries or illnesses who seek to return to driving. It also makes it important for those with chronic ongoing conditions whose ability to drive safely needs to be determined by health professionals. Clinical and neuropsychological examinations and assessment criteria for a variety of chronic diseases, injuries or disabilities have been previously discussed in the literature including Alzheimer disease, Parkinson disease, cardiovascular accidents, traumatic brain injuries, sleep apnea and narcolepsy to name a few (Marino *et al.* 2013). However, in many cases, there is still little knowledge about the underlying brain mechanism of drivers with these conditions in addition to the lack of consensus on many clinical evaluations of driving ability (Hung *et al.* 2014). As a result, although in a limited way, a number of neuroimaging studies have addressed the issue of fitness to drive for a range of underlying neurological impairments including sleep apnea (Risser and Ware 1999), epilepsy (Yang *et al.* 2010, Krestel *et al.* 2011) and brain lesions (Papageorgiou *et al.* 2012, Hung *et al.* 2014).

Previous studies based on retrospective surveys of driving and accident experience and self-completion questionnaires have shown evidence of increased risk of severe traffic accidents in drivers with a history of epilepsy (Taylor *et al.* 1996). The study of Yang *et al.* (2010) investigated the loss of consciousness in patients with epilepsy using virtual-reality driving simulation while monitoring subjects using EEG. They examined the hypothesis that seizure types involving a greater loss of consciousness are most likely to cause collisions and that different types of seizures have different effects on driving ability. Their findings demonstrated that impairment in driving performance during seizures, in terms of magnitude and character, differed by type of seizure. Krestel *et al.* (2011) analysed the impact of interictal epileptic activity, without clinically overt seizure symptoms, on driver reaction time while collecting EEG signal. They observed that epilepsy patients had slower reaction times during generalised interictal epileptic activity compared to reaction times during unremarkable EEG periods. They suggested reaction-time EEG as a possible tool to assess driving ability.

Cerebral control plays an important part in driving due to its involvement in motor control and movement planning, action preparation, and monitoring other vehicles and maintaining a safe distance. On the other hand, cerebral damage is a common occurrence due to infection, neural degeneration and physical trauma. To understand neural mechanisms and brain plasticity about human driving behaviour and strategies of assessment and rehabilitation for drivers with neurological conditions, Marino *et al.* (2013) used fMRI virtual-reality technology to assess driving ability after brain damage by experimenting with both normal subjects and clinical subjects with focal cerebral damage. Their findings showed that drivers with cerebral damage showed significantly poorer speed control. Papageorgiou *et al.* (2012) considered the case of patients with homonymous visual field defects due to unilateral vascular brain lesions in a collision-avoidance task and MRI paradigm. They observed no significant difference between patients with left-hemisphere and right-hemisphere lesions among their findings. Details of these studies are summarised in Appendix C.

## 4. *Brain activity of drivers in semi-automated settings*

With semi-automated or conditionally automated driving getting ever closer to becoming a reality on the roads, research into the behavioural aspects of driving in such settings as well as the interaction of drivers with semi-automated cars has surged (Sportillo *et al.* 2018). Topics such as driver takeover and vehicle disengagement are heavily studied in mainstream road safety research (Favarò *et al.* 2018, Yoon *et al.* 2019). Neuroimaging driving behaviour studies have weighed into this problem as well although the existing efforts in this domain are limited.

Lee and Yang (2020) considered the case of human-machine interaction associated with Level 3 automated vehicles in which the takeover of the human driver from the system in an event of an emergency is an inevitable element driving. They tested a range of takeover transition alerts while analysing drivers' EEG brainwaves. The transition alert system that used a combination of auditory, visual and haptic stimuli proved to be most effective. Tsunashima and Yanagisawa (2009) used fNIRS to evaluate the brain function of car drivers with and without Adaptive Cruise Control (ACC) systems and demonstrated that the frontal lobe was less active during ACC drive. Cao *et al.* (2019) conducted an attention-task simulated driving experiment in which drivers were introduced to compensate the shift and steer the car back to the centre of the cruising lane in response to unexpected random lane-departure events. They showed how EEG signal can be used as a measure of driver's attention to these random takeover tasks. Arakawa *et al.* (2019) studied the psychological assessment of driver mental state in autonomous vehicles using fNRIS signal in combination with other biomarkers such as blood pressure, body pressure distribution, salivary monitoring and eye tracking. Details of these studies are summarised in Appendix D.

## 5. *Brain activity of fatigued/drowsy drivers*

Driver fatigue – generally manifesting as drowsiness, tiredness, weakness and lack of alertness and causing deteriorated vigilance, risk of driving errors and poor decision making (Barua *et al.* 2019) – has been established as a major cause of road accidents and has been implicated in 20-30% of road fatalities (Lal *et al.* 2003, Kar *et al.* 2010). A significant portion of road accidents occurring at night-time can be attributed to this issue. As a result of the high prevalence of fatigue-related traffic accidents, much research in the road safety domain has been devoted to the detection and quantification of fatigue, to determine indicators and measures of driver sleepiness (Jagannath and Balasubramanian 2014, Al-libawy *et al.* 2018, Wei *et al.* 2018). The majority of these efforts have been aimed at developing Automated Driver Sleepiness Detection (ADSD) systems (Kong *et al.* 2017, Wang *et al.* 2017, Jacobé de Naurois *et al.* 2018, McDonald *et al.* 2018, Jacobé de Naurois *et al.* 2019). In more recent years, this line of research has also been extended to the case of (semi)-automated driving settings in which the driver's role may change to an active operator to a fallback-ready driver, thus intensifying the monotonicity of the drive and increasing the possibility of driver sleepiness and lack of attentional availability when the take-over of control is required (Naujoks *et al.* 2018, Vogelpohl *et al.* 2019, Wu *et al.* 2019).

In addition to the subjective ratings and self-reported questionnaire-based measures of driver fatigue (Ferguson *et al.* 2012), indicators of driver sleepiness can be classified as (i) vehicle-based performance measures, such as variability in lateral vehicle control, or safe distance to other vehicles, (ii) driver-based behavioural measures, such as ocular parameters and saccadic movement, eye closure, blink duration, and yawning, and (iii) physiological measures, such as respiration, heartrate, and EEG, as well as combinations of these indicators (Kar *et al.* 2010, Barua *et al.* 2019).

Among physiological indicators, EEG has been the most commonly used metric in automatic drowsiness detection of drivers. It has also been suggested to be the most reliable indicator of fatigue and sleepiness detection for drivers. Increase alpha band power during driving in particularly has been associated with drowsiness (Simon *et al.* 2011). Various data-driven approaches have so far been adopted in order to develop EEG-based driver drowsiness detection systems, including single and multi-scale (i.e. ensemble) entropy measures (Kar *et al.* 2010, Huang *et al.* 2015, Hu 2017a, Hu 2017b, Fonseca *et al.* 2018, Hu and Min 2018), a variety of machine learning methods (Ma *et al.* 2019) and methods such as Support Vector Machine (SVM) (Barua *et al.* 2019, Chen *et al.* 2019), Wavelet Transformation (Kar *et al.* 2010), Convolutional Neural Network (CNN) (Gao *et al.* 2019), k-means clustering (Gurudath and Riley 2014) and logistic regression (Babaeian *et al.* 2016).

Subjective measures of self-reported sleepiness, predominantly Karolinska Sleepiness Scale (KSS) (Åkerstedt and Gillberg 1990), have been the most common measure as the ground truth of the classification methods in these studies (Kar *et al.* 2010), although this measure has often been supplemented by other physiological or behavioural indicators as well (Zhao *et al.* 2011, Morales *et al.* 2017). While the predominant method of experimentation in this category of studies has been simulated driving, a limited number of studies have tested ADSD system applications in real-world field driving settings (Perrier *et al.* 2016, He *et al.* 2018, Chen *et al.* 2019). In the majority of studies in this category, sleepiness is involved by sustained and monotonous driving in the simulated (or field) setting. But some experiments have conducted the drowsy condition experiments in late night or after midnight sessions and with real sleep-deprived subjects (Perrier *et al.* 2016, Ahlström *et al.* 2019, Barua *et al.* 2019). A summary of these studies is available in Appendix E.

## 6. *Brain activity of drivers during decision-making tasks*

Neuroimaging studies of driving behaviour have addressed a number of distinct dimensions related to decision-making aspects of driving. Particular dimensions are driver uncertainty perception, risk-taking behaviour and their elation with peer influence as problems that have also been investigated in mainstream road safety research (Jonah 1986, Møller and Gregersen 2008, Brandau *et al.* 2011, Møller and Haustein 2014, Abay and Mannering 2016, Mirman and Curry 2016). Another notable topic addressed by neuroimaging studies is related to the prediction or anticipation of driver decisions such as braking, acceleration or turning (Kim *et al.* 2014, Zhang *et al.* 2015, Vecchiato *et al.* 2019), with this topic being virtually exclusive to the brain imaging sector of road safety research.

As pointed out in the introduction of this article, there is an abundance of studies on neural correlates of decision-making particularly for those decisions with economic relevance (Sanfey *et al.* 2003, Braeutigam 2005, Kenning and Plassmann 2005, Sanfey *et al.* 2006, Camerer 2008,

Clithero *et al.* 2008, Livet 2009, Rustichini 2009, Smith and Huettel 2010), yet relatively little on neural correlates of decision-making in driving. In a pioneering neuroimaging driving behaviour study, Callan *et al.* (2009) point out this gap, while also making a distinction between reward-weighted and cost-weighted mechanisms of decision-making and pointing out that the latter has been far less represented in neuroimaging decision studies. They also argued that given the risks involved in driving, the decision-making behaviour of drivers is mostly cost-weighting in nature. They studied neural correlates of resolving uncertainty in drivers' decision-making by simulating turning-right scenarios in left-hand traffic at a signalised intersection while collecting fMRI signal. The uncertainty was created by a big truck occluding drivers' view. They observed that resolving uncertainty resulted in a reduced activity in the amygdala and anterior cingulate and concluded that these areas may be implicated in cost-weighted decision-making.

Tanida *et al.* (2018) also made a distinction between the anticipatory and the surprise model of car driving, as a goal-directed behaviour, consistent with the model originally proposed by Tanida and Pöppel (2006). In their own terms, they argued that "An anticipatory mode of car driving, which is based on future directed anticipations and accordingly positive feedback signals, might be related to perceived safety, while a surprise mode of driving might be related to perceived risk." (Tanida *et al.* 2018) (p. 109). In their study, they used a controlled introspection technique, a mental imagery technique, and memory retrieval of risky versus safe driving episodes while also collecting fMRI brain activity. Their observation suggested that perceived safety is associated with a higher involvement of visual and motor areas and the left anterior cingulate cortex (ACC).

The studies of Chein *et al.* (2011) and Vorobyev *et al.* (2015) introduced the role of peer influence on adolescent risk-taking behaviour – as a relatively well-studied road safety topic (Smorti *et al.* 2014, Weston and Hellier 2018) – to the neuroimaging driving experiments. Chein *et al.* (2011) measured brain activity in adolescents, young adults and adults as they made simulated driving decisions while subjects being under the impression that their decisions are observed by their peers in certain scenarios (blocks). The study showed that adolescents' brain activity was greater in reward-related brain regions in peer observation blocks, which explains the increase in risk-taking tendencies of adolescents in the presence of peers. The study of Vorobyev *et al.* (2015) similarly suggested an enhanced reward-processing of risk taking in adolescents under peer influence, in that, "a decision to take a risk activated the adolescent brain much more than a decision to stay safe" (p. 16). Sub-groups of high and low risk takers, determined based on personality tests, did not show different patterns of brain activation. But when defined based on the actual (observed) risk-taking performance, between group differences in terms of brain activation was observed. High-risk-taker adolescents showed less strong activity in two areas of the left medial prefrontal cortex compared to low-risk-takers during risky decision-making tasks. They concluded that this shows an increased cognitive effort to take risks in low-risk-taker adolescents during risky decisions.

A different theme of studies in the neuro-ergonomic and road safety domain has focused on the anticipatory recognition of driver emergency actions that could potentially be used in the development of driving assistance systems. Such systems could assist in the early preparation of the vehicle for emergency responses that can compensate for driver reaction time. This stream of work has been conducted based on the premise that brake pedal deflection is preceded by cognitive processes that are, to certain degrees, observable in the central nervous system and brain activity. It, therefore, utilises the possibility of using brain signals that precede the moment the brake pedal is activated to perform an early detection and preparation of emergency braking (Hernández *et al.*

2018). This cohort of studies has focused on predicting specific drivers' actions such as braking and accelerating and even turning using EEG signal (Zhang *et al.* 2015). Haufe *et al.* (2011) used EEG potentials to predict upcoming emergency braking in simulated driving experiments. They demonstrated how cerebral activity can be used as an indication of emergency braking prior to a behavioural response. The authors subsequently replicated this methodology during naturalistic driving on a non-public test track and observed similar behavioural patterns and neurophysiological phenomena in the field test setting when compared to the simulated driving counterparts (Haufe *et al.* 2014). This methodology was further extended to enable the researchers to distinguish between different types of emergency braking, e.g. sharp braking and soft braking, in diverse traffic conditions using a feature combination approach (Kim *et al.* 2014). Hernández *et al.* (2018) used Support Vector Machine (SVM) and Convolutional Neural Networks (CNN) to differentiate EEG signals associated with braking intention and that of normal driving, under various cognitive states such as mental workload, stress and fatigue. These studies have collectively identified neurophysiological patterns of sensory perception and processing that characterised driver emergency braking prior to the action taking place. For example, it has been established that dorso-mesial premotor cortex has involvement in the preparation of foot movement for braking and acceleration actions (Vecchiato *et al.* 2019). Details of these studies have been summarised in Appendix F.

## 7. *Brain activity during general driving tasks*

Apart from the neuroimaging studies on the brain activity of intoxicated drivers, distracted drivers, drowsy and fatigued drivers, drivers with neurological conditions, as well as the those related to specific decision-making tasks, certain neuroimaging studies have considered the general neural correlates of driving by conducting brain imaging experiments of general driving activity (e.g. steering, car following, maintaining safe distance) involving no secondary task with healthy well-rested drivers. These studies and their findings are reviewed in this section.

The fMRI studies in this category have collectively provided great insight into the brain areas, circuits and networks that are active during driving. The majority of these findings have been obtained from measuring the BOLD response of drivers and comparing them across passive and active driving blocks. The pioneering fMRI study of Walter *et al.* (2001) showed that driving requires the coordinated activity of occipito-parietal and motor areas of the brain. Calhoun *et al.* (2002) investigated the modulation of brain activity with driving speed. They observed the correlation between driving speed and activity in areas associated with error monitoring and inhibition as well as areas associated with vigilance. Activity in these areas showed a decrease at higher driving speeds. Uchiyama *et al.* (2003) investigated neural substrates of driving at a safe distance and observed increased cerebellum activations indicating its role in providing visual feedback during tracking of the front car. Driver performance in maintaining a constant headway was also observed to be correlated with activity in anterior cingulate, which is a reflection of its role in error detection. Spiers and Maguire (2007) identified brain regions that are associated with prepared actions such as turning, reversing and stopping as opposed to brain regions associated with unexpected events such as avoiding collisions. Mader *et al.* (2009) considered the role of route familiarity on driving attention process and investigated cerebral activation patterns of driving on a familiar versus an unfamiliar route. Overall, these fMRI studies have identified a network of brain areas that underlie the cognitive mechanism of driving including a large core (or common) circuit (Navarro *et al.* 2018) involving cerebellum (responsible for motor coordination), the bilateral extrastriate cortex (involved in visual attention), the right middle temporal gyrus (linked to visual motion processing), bilateral precuneus (involved in motor coordination), the left anterior part of the insula (associated with representation subjective feelings), the right posterior cingulate gyrus (implicated in topographical memory) and the right dorsomedial part of the thalamus (which contributes to perception, attention and timing) (Navarro *et al.* 2018).

Compared to the aforementioned fMRI studies, a more recent cluster of EEG experiments (both in lab and in field settings (Protzak and Gramann 2018)) have also been reported in this category within the last few years. Garcia *et al.* (2017), for example, differentiated between proactive and reactive brain states during driving. The former is responsible for actively planning a response according to sensory information and is characterised by delta-beta activity. The latter is responsible for processing incoming information and is characterised mainly by activity in the alpha band. The EEG method has also been applied to classify various patterns of brain activity associated with different personalities and styles of driving (Yang *et al.* 2018b, Ding *et al.* 2019, Yan *et al.* 2019a, Yan *et al.* 2019b) as well as different levels of driver's workload (Di Flumeri *et al.* 2018, Foy and Chapman 2018). The methodology proposed to infer driver's mental workload from their EEG brain activity has also been tested and validated via field studies (Di Flumeri *et al.* 2018). The findings of the studies of this category have been synthesised in Appendix G.

## 8. Summary statistics

In the previous section, the core data set of 85 journal articles on applications of neuroimaging methods in driving research was reviewed and synthesised. Seven different categories of these studies were identified and surveyed in that section. Figure 5 illustrates a purely selective number of sample images from these studies on applications of the fMRI, MEG, EEG or fNIRS methods, either in a simulated or on-road setting, in experiments of driving behaviour, as surveyed in this work. Figure 6 provides an abstract visualisation of these studies, their underlying brain imaging method and the category of research theme to which they belong. Figure 7 shows the temporal evolution of journal articles in the core data set within each category as well as that of the total number of neuroimaging driving behaviour studies.

According to these graphics, the focus of the studies that have adopted neuroimaging methods in driving behaviour research has been unevenly distributed across various topics within this domain. Also, these methods have different degrees of representation across different topics or themes of study. Overall, applications of brain imaging methods have been dominated by fMRI and EEG studies, while applications of fNIRS and MEG have been very limited. Studies with a focus on fatigue/drowsy driving, distracted driving and general driving tasks appear to have constituted the majority of the body of neuroimaging applications in driving behaviour research. On the other hand, and by contrast, very few applications of these methods to the experimentation of driving in semi-automated settings or drivers with neurological conditions have been reported. While brain imaging studies on intoxicated and distracted drivers have been dominated by fMRI applications, in studies of driving under fatigue/drowsy condition, EEG appears to have been the exclusive method of choice. In terms of the rate of publication, based on Figure 7 it appears that journal articles have been appearing at a greater rate since 2016. This is to certain degrees consistent with the data presented in Figure 1 which is based on a broader and more inclusive set of studies (e.g. conference papers, editorial etc) that suggested an elevated attention to this domain since 2014.

In terms of the temporal evolution of neuroimaging in driving behaviour research, and again based on the core data set presented in Figure 7, it appears that at early stages of the development, the research had been more focused on understanding the neurological bases of driving in tasks and under no adverse stimulus. These efforts were then followed by more studies on neural correlates of intoxicated and distracted driving that have both been well represented in the literature of this domain since 2007. Another interesting aspect, that was also reflected in a clear way in the macro-scale analyses of the broader literature presented in Section 3, is the evident spike in EEG studies of fatigue/drowsy driving and research attempts towards the development of automated driver fatigue detection systems based on brain signals. Currently, in terms of the rate of publications, this appears to be the hottest topic of research in this domain. And as pointed out in Section 4.4, Figure 7 also indicates that the attention to neuro-cognitive aspects of semi-automated driving has been generated only within the last few years with the topic being clearly underrepresented in this literature and by far the least studied topic. Studies of brain imaging related to the brain activity of non-healthy drivers suffering from brain injuries or chronic neurological conditions appear to be the second least studied topic thus far.

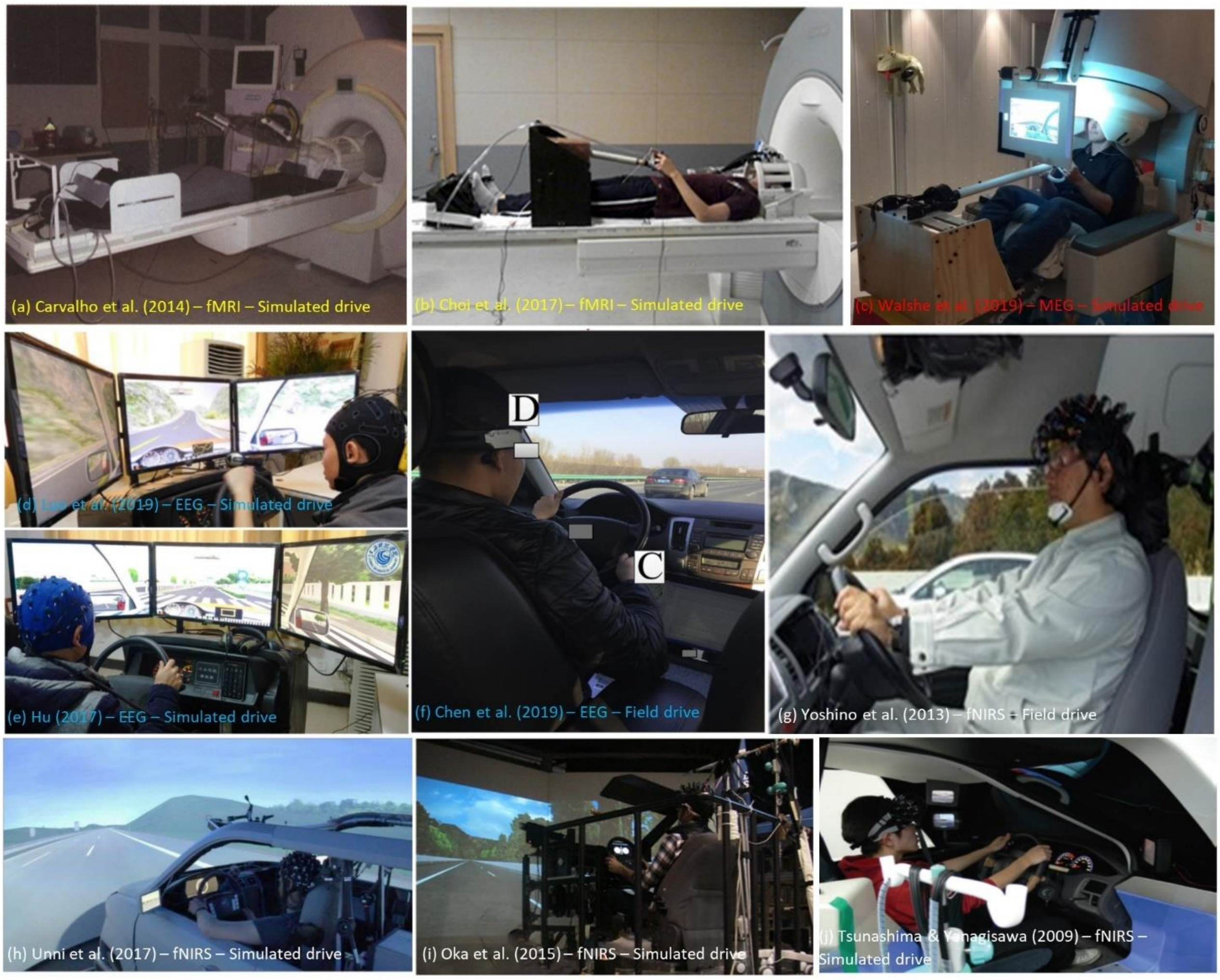

**Figure 5** Samples of images from the published studies on neuroimaging driving behaviour experiments, in driving simulation or field (i.e. on-road) settings, based on each of the four methods surveyed in this review: fMRI, MEG, EEG and fNIRS. The number of sample images associated with each method is not proportional to the prevalence of that method. The images have been borrowed from studies of (a) Carvalho *et al.* (2014), (b) Choi *et al.* (2017), (c) Walshe *et al.* (2018), (d) Luo *et al.* (2019), (e) Hu (2017b), (f) Chen *et al.* (2019), (g) Yoshino *et al.* (2013), (h) Unni *et al.* (2017), (i) Oka *et al.* (2015), (j) Tsunashima and Yanagisawa (2009).

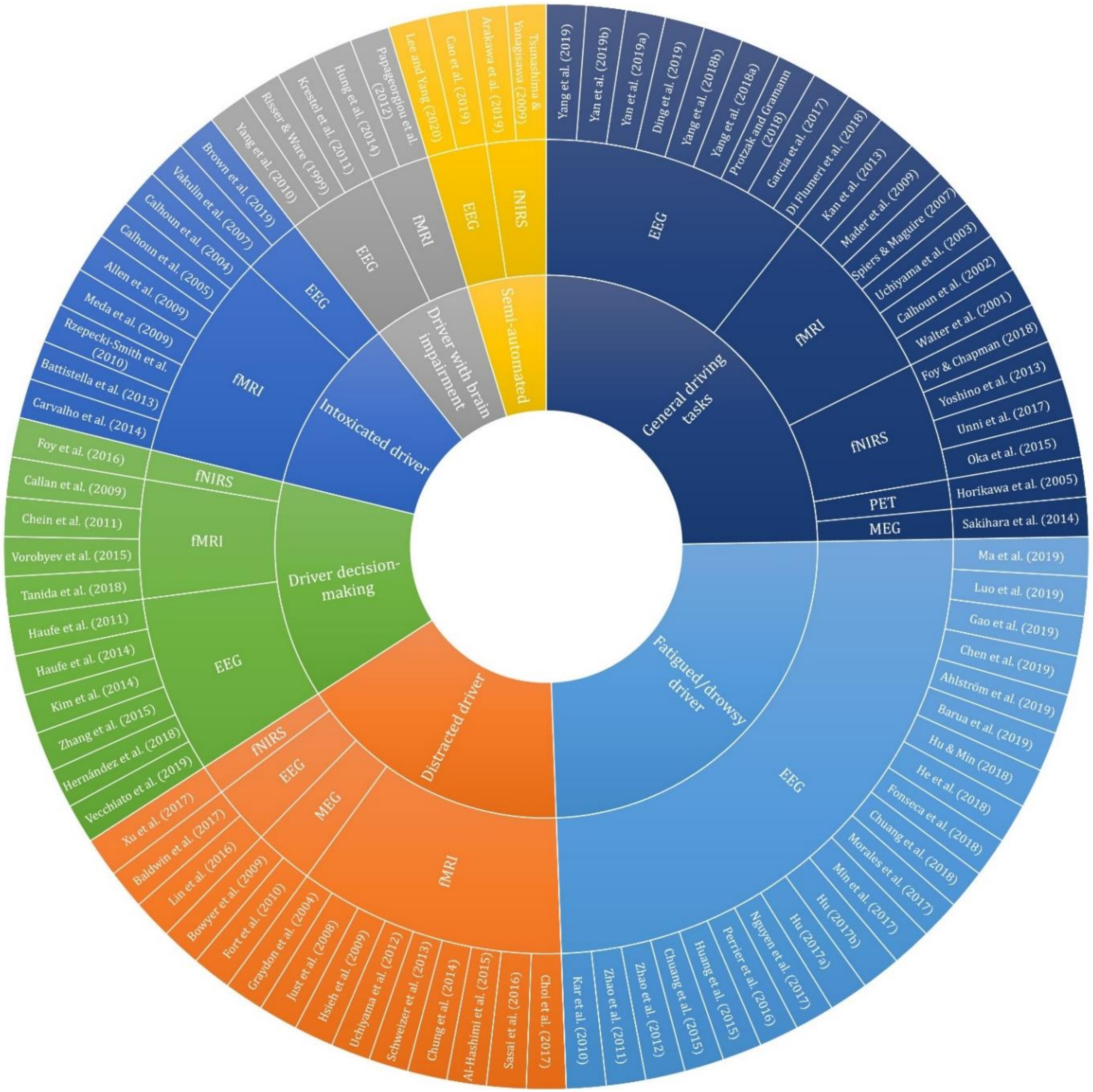

**Figure 6** The core set of surveyed studies on brain activity during driving tasks, along with their method of brain imaging and their general the

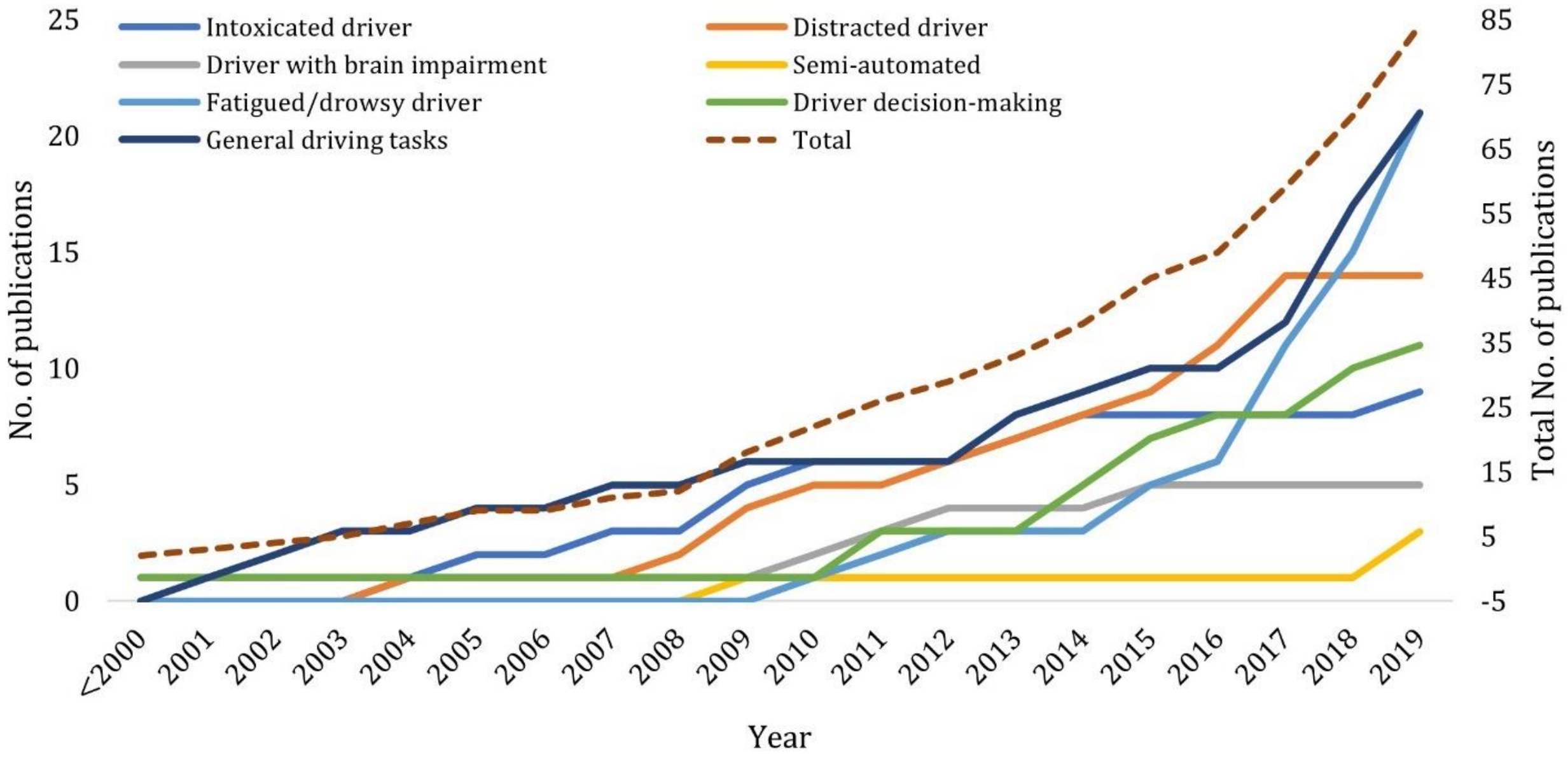

**Figure 7** The temporal trend of the number of publications on each of the seven themes of the brain imaging studies of the driving behaviour according to the core reference database surveyed in this review. The dashed line shows the total number of publications in all seven categories (themes) and is represented by the right vertical axis. The data related to 2020 only reflects the first three months, and not the full year.

## 9. Discussions, conclusions and directions for future research

Driving as a complex behavioural mechanism relies on the integration of information and concurrent execution of various behavioural functions that allow drivers to maintain situational awareness as well as visual and auditory vigilance, manage distractive stimuli, make appropriate decisions and often respond to unexcepted events within a short period of time. As illustrated in this work, applications of brain imaging methods have enabled us to acquire a clearer picture of the neural components and cognitive mechanisms of driving. This stream of research not only has informed us of the general neural activity of healthy drivers in traditional driving settings under sober, well-rested and non-distracted conditions (Navarro *et al.* 2018), but also has allowed us to understand the mechanisms by which driver performance is compromised by external elements such as the consumption recreational drugs (Calhoun *et al.* 2004a, Calhoun and Pearlson 2012) or engagement in secondary tasks (Bowyer *et al.* 2009, Palmiero *et al.* 2019). As a result of the experimental efforts in functional imaging of driver brain, much more is known about the common brain areas and networks in a healthy brain that are active during driving as well as the areas and circuits whose functions are disrupted as a result of such adverse stimuli. In addition, thanks to the wealth of evidence obtained predominantly from EEG studies, typical patterns of electrical brain activity of drivers during the fatigue or sleepiness state (Chuang *et al.* 2015) or the typical patterns that precede critical actions such as emergency braking (Haufe *et al.* 2014) have been identified and have been utilised for the development of automatic detection/prediction systems that can enhance the safety of driving in various ways.

As evident by this critical review of the literature, various neuroimaging methods each offer certain possibilities, advantages and limitations for studies related to neuro-ergonomics of driving. Issues such as cost of operation, mobility of equipment, degree of invasiveness of experiments, confinement of subjects during tasks, preparation time for experiments, sensitivity of data to subject's body movement, and the temporal and spatial resolution of the collected brain signal and the interplay between these factors make certain brain imaging methods more suitable for certain research questions/applications in this context. Among the four main methods that are used for brain data collection in driving behaviour studies, EEG and fNIRS have mobile equipment whereas fMRI and MEG experiments require fixed scanners that are not portable. Experiments of fMRI and MEG also require that the participant's head be confined in a small space and their data are highly sensitive to the head movement compared to EEG and fNIRS. These characteristics automatically make applications of MEG and fMRI an impossibility for field driving experiments. These technical limitations also often require that the experiments of simulated driving be conducted using more simplified computer interfaces and visualisations as opposed to EEG and fNIRS equipment that can be integrated with highly sophisticated simulators and even on-road driving tests. Although much advancements have been made to represent all relevant components of driving in fMRI experimental settings (See Figures 5(a) and 5(b) as prime examples of such state of the practice technological developments), issues of ecological validity and subject immobility remain as restrictive features of the fMRI studies (and MEG studies to a lesser extent) in a driving context compared to EEG and fNIRS methods of brain data collection. This is in addition to the fact that fMRI and MEG studies are expensive experiments due to the specialty facilities and equipment that they require and the high operational costs. On the other hand, MEG and fMRI data have a very high spatial resolution (in the order of millimetres), thus suitable for more accurate localisation of various brain functions during driving, as opposed to the EEG method for example that has a spatial resolution of centimetres. EEG signal, however, has a high temporal resolution and it can accurately capture changes in brain electrical activity that occur quickly. Poor temporal resolution is a clear disadvantage of fMRI, as a method that offers a delayed representation of cortical activity. EEG is also a fairly and viably non-invasive and inexpensive method compared to the other alternative methods. It should also be noted that although fNIRS equipment is technically portable and although it does not impose strict head movement constraint, in terms of the invasiveness, it is deemed as a relatively uncomfortable method to participants given that it requires attachment of probes on the scalp. But, since it does not expose subjects to magnetic fields, it can accommodate participants with ferromagnetic implants without any safety concern.

As evidenced by the current survey of the literature, applications of the neuroimaging methods in driving behaviour research have been relatively diverse and the existing studies have encompassed a large variety of research questions in this domain. However, certain topics have been relatively less represented while it appears that the neuroimaging methods could offer unique possibilities in those areas. For example, the vast majority of studies have been conducted by relatively young and healthy drivers. Whereas, the studies of aging brain or drivers recovering from brain injuries or stroke or those suffering from chronic neurological illnesses – all issues that are critical for the accurate assessment of fitness to drive (Marino *et al.* 2013, Devos *et al.* 2015) – are clearly underrepresented in this domain. Given the increases in life expectancy in many societies and the desire of people to maintain their license to drive or re-gain them post injuries or despite chronic conditions, these problems are currently of relevance. It is believed that there are many areas to be explored using neuroimaging techniques in this sub-domain of driving neuro-ergonomics literature. Also, contrasts of driver brain activity across adolescents, adult and older drivers could offer greater insight into the fundamental behavioural differences of these sub-groups during driving and provide explanations as to phenomena such as teen drivers being more prone to fatal crashes or risky behaviour (Freydier *et al.* 2014, Walshe *et al.* 2017). The relative risk of drivers with AD/HD or depression compared to healthy drivers has long been a matter of debate in the road safety literature (Reimer *et al.* 2010, Vaa 2014, Aduen *et al.* 2015, Aduen *et al.* 2018). It appears that brain imaging methods could shed more light on these questions too and provide some explanations as to the mechanisms of brain activity in AD/HD drivers for example that make them more susceptible to unsafe driving. Neuroimaging studies of driving under the effect of recreational substances were well represented in our survey of the literature. But it should be noted that these studies have predominantly focused on the effects of alcohol intoxication while there are only a few equivalent studies related to cannabis intoxication (Battistella *et al.* 2013, Brown *et al.* 2019). As a result, while relatively much is known about the neuro-cognitive impacts of alcohol intoxication of driving performance, the equivalent knowledge in relation to the cannabis effect is yet to be developed. Given the increasing rate of legalisation of cannabis around the world and the need for evidence-based regulations in the context of driving (Bergeron and Paquette 2014), it could be timely and relevant for neuroimaging experiments to further explore this question. Also, while neural correlates of distracted driving constitute a great portion of the studies in this domain, there are more nuanced questions in relation to neuro-cognitive effects of having conversations while driving that could be investigated by future studies. These nuances include issues such as underlying differences in brain activity modulated by the sentiment of the conversation or whether the conversation is in a person's native or second language. The emergence of semi-automated vehicles also clearly calls for more attention from scientists with neuroimaging expertise in this domain as issues such as drowsiness, mind-wandering and ability to take back control when required are still underexplored in semi-automated driving settings.

## Appendix A. Summary of studies on brain activity of intoxicated drivers

| Reference | Signal type | Subject tasks | Substance | Design factors [design type] | Subjects | Behavioural measures | Brain imaging findings | Behavioural findings |
|---|---|---|---|---|---|---|---|---|
| Calhoun *et al.* (2004a) | fMRI | Active & passive driving | Alcohol | Alcohol dose (BAC 0.04, 0.08, placebo) [within subject] | 9 | Collisions; Near collisions; Lane deviations (number, duration); Speed limit violation | Identifying seven different driving brain networks with different time courses; Identifying does-related changes in fMRI signal in orbitofrontal and motor regions | Disruptive, dose-dependent effects of alcohol on several brain circuits; Compared to sober, at the lower BAC, performance slightly improved and subjects reduced average speed; At the higher BAC, subjects drove at higher speeds and collisions increased |
| Calhoun *et al.* (2005) | fMRI | Active & passive driving | Alcohol | Alcohol dose (0.04, 0.08, placebo) [within subject] | 10 | Collisions Near collisions Lane deviations (number, duration); Speed limit violation | Specific brain circuits differentiated modulated by alcohol; No global decrease in brain activity due to alcohol; Localised increase/decrease identified; ICA revealed a set of driving-related networks; Significant alcohol related changes found in OF/anterior cingulate cortex and primary motor cortex | Disruptive, dose-dependent effects of alcohol on several brain circuits; Alcohol consumption impaired error detection region; Significant increase in collision and near collision at higher BAC; Less responsive to peripheral events & depressed perceptual and motor functioning of inebriated drivers |
| Vakulin *et al.* (2007) | EEG | Sustained simulated drive | Alcohol | Alcohol alone, Sleep deprivation alone & Combination [within subject] | 21 | Steering deviation; Braking reaction time; Collisions | Combination of sleep and alcohol condition resulted in a significant increase in alpha/theta EEG activity | Performance measures significantly affected by time; Steering deviation significantly increased in the combined alcohol and sleep condition |
| Allen *et al.* (2009) | fMRI | Driving while performing a secondary (visual oddball) task | Alcohol | Alcohol dose (moderate, high, placebo) [within subject] | 40 | Response time; Correct hits; False alarms; Line crossings; Speed; Braking; Steering weave | Dose-dependent linear decrease of fMRI signal in hippocampus, and anterior cingulate | Dose-dependent linear increase in reaction time; No effects associated with correct hits or false alarms; Lesser speed decrease due to secondary task, at high alcohol dose; Passenger-side line crossings significantly increased at high dose; Better driving performance in moderate dose condition; The ability to perform a secondary task is |
| | | | | | | | | impaired at BAC levels above the legal limit |
| Meda *et al.* (2009) | fMRI | Active & passive driving | Alcohol | Alcohol dose (0.05%, 0.10%, 0.0%) [between subject] | 40 | Line crossings; Speed; Crashes; Steering weave | The high dose alcohol caused significant impairment of brain functionality related to motor planning, control, error monitoring and memory | Significant does-dependent changes in line crossing and mean speed |
| Rzepecki-Smith *et al.* (2010) | fMRI | Active & passive driving | Alcohol | Alcohol dose (BAC 0.10, placebo) [within subject] | 40 | Line crossings; Speed; Collisions; Steering weave | Specific disruptions of functional connectivity between the frontal-temporal-basal ganglia and the cerebellar circuits at the active dose | Unstable motor vehicle steering at the active dose |
| Battistella *et al.* (2013) | fMRI | Active & passive visual tracking | Cannabis | Cannabis dose [within subject] | 31 | Psychomotor skills | Cannabis altered the network involved in saliency detection; Cannabis increased self-oriented mental activity; Cannabis altered the activity of the brain networks involved in cognition (even at low dose) | Cannabis impaired psychomotor skills |
| Carvalho *et al.* (2014) | fMRI | Fixation, active driving & passive driving | Alcohol | Alcohol dose (0.04%, 0.08%, 0.0%) [within subject] | 9 | Steering; Pedal activity; Speed | Cerebellar areas showed signal increases (decrease) during steering maintenance (steering changes); Activation in motor areas increased during braking | Compared with sober baseline, at the lower BAC, behavioural performance slightly improved and participants reduced average speed; At the higher BAC, subjects drove at a higher average speed |
| Brown *et al.* (2019) | EEG | Drive in various environments | Cannabis | Cannabis dose (6.7 THC, placebo) [within subject] | 11 | Lane position | Significant relation between impaired driving performance and EEG power in slow theta band in parietal and occipital areas; EEG biomarkers correlated with SDLP | Heart rate was significantly higher during driving in dosed sessions compared to placebo |

## Appendix B. Summary of studies on brain activity of distracted drivers

| Reference | Signal type | Subject tasks | Distractive stimulus | Design factors [design type] | Subjects | Behavioural measures | Brain imaging findings | Behavioural findings |
|---|---|---|---|---|---|---|---|---|
| Graydon *et al.* (2004) | fMRI | Passive drive | Vidual event detection | Visual detection only, drive only, dual task [within subject] | 6 | N.A. | Visual event detection during driving engages multiple cortical and sub-cortical neural systems that are interconnected and work in concert | N.A. |
| Just *et al.* (2008) | fMRI | Drive on a curvy road | Sentence comprehension | Drive alone vs drive while listening [within subject] | 29 | Driving errors; Path deviation | The parietal lobe activation associated with spatial processing decreased by 37% when participants concurrently listened to sentences | Processing of the auditory sentences resulted in a significant deterioration in driving accuracy |
| Bowyer *et al.* (2009) | MEG | Visual detection while driving passively | Covert conversation | Drive with and without conversation [within subject] | 24 | Reaction time | Reaction times inversely related to the amount of brain activity detected in the right superior parietal lobe | Reaction time for the conversation task slightly longer than that of the baseline |
| Hsieh *et al.* (2009) | fMRI | Visual detection in passive driving | Covert conversation | Long, short and no conversation [within subject] | 10 | Reaction time; Miss rates | Increased brain activation under the conversation condition in language areas in addition to other areas | Covert conversation resulted in longer visual event reaction times compared to driving with no conversation; The effect of miss rate was negligible |
| Fort *et al.* (2010) | MEG | Drive while reacting to visual events | Listening to radio | Level of attentional demand (Simple vs dual task) [within subject] | 13 | Reaction time | Increased attentional demand affects neuronal processing of visual information, even at perceptual stage | Time required for making the decision was significantly different between the simple and dual task |
| Uchiyama *et al.* (2012) | fMRI | Car following | Sentence comprehension & tone discrimination | Baseline passive, drive only, auditory only, dual task [within subject] | 18 | Following; distance | Decline of brain activity in medial prefrontal cortex and right inferior parietal lobe may explain poorer car-following performance | Car-following performance was worse during the dual task than during the single-driving task |
| Schweizer *et al.* (2013) | fMRI | Drive straight and make turns | Sentence comprehension | Oncoming traffic & auditory distraction [between subject] | 16 | Lane position; Speed | During distracted driving, brain activation shifted dramatically from the posterior, visual and spatial areas to the prefrontal cortex | Lane position during straight driving was not significantly different from that of distracted straight driving |
| Chung *et al.* (2014) | fMRI | Drive straight | Arithmetic calculation | Drive only & drive distracted | 16 | N.A. | Regions responsible for error monitoring and control of unnecessary | N.A. |
| | | | | [between subject] | | | movement showed increased activation during distracted driving compared with driving only | |
| Al-Hashimi *et al.* (2015) | fMRI | Active and passive drive while responding to sign stimuli | Vidual event detection | Drive only, sign only, auto pilot and sign, drive and sign [within subject] | 31 | Reaction time | Only a single brain region, the superior parietal lobule, exhibited a significant relationship with multitasking performance | Significantly slower responses during the sign and drive condition compared to the counterpart auto-drive |
| Lin *et al.* (2016) | EEG | Correct induced lane departures | Mind wandering | Visual only vs visual & motion feedback [within subject] | 10 | Reaction time | In the absence of salient sensory information, activation in the frontal-parietal attention network is stronger | Mind-wandering during driving tends to occur when perceptual demand is low |
| Sasai *et al.* (2016) | fMRI | Drive & change lane | General auditory stimulus | Nature of auditory distraction (GPS vs radio show) [within subject] | 13 | Drowsiness; Lane deviation | The brain may functionally split into two separate “driving” and “listening” systems when the listening task is unrelated to concurrent driving, but not when the two systems are related | No significant difference was found in behavioural measures between integrated and split task conditions |
| Baldwin *et al.* (2017) | EEG | Monotonous drive | Mind wandering | N.A. [within subject] | 9 | Speed; Lane deviation | Periods of mind wandering were associated with increased power in EEG alpha band | Self-reported mind wandering frequency was high during driving, and did not statistically change over days of participation |
| Choi *et al.* (2017) | fMRI | Drive while performing sub-tasks | Carry out calculations | Passive drive, drive alone, drive with sub-task [within subject] | 15 | N.A. | The number of activation voxels greatly decreased in the parietal area under drive with sub-task condition; Task-performing areas, such as the inferior frontal gyrus and the superior temporal gyrus, showed increased activation | N.A. |
| Xu *et al.* (2017) | fNIRS | Drive straight | Auditory and vidual task | Drive only, drive while distracted [within subject] | 12 | Speed; Driving error | Secondary tasks during driving led to brain activity changes, and dynamic configuration of the connectivity | Secondary tasks during driving led to poor driving performance |

## Appendix C. Summary of studies on drivers with brain impairment and underlying neurological illness

| Reference | Signal type | Subject tasks | Damage type | Design factors [design type] | Subjects | Behavioural measures | Imaging findings | Behavioural findings |
|---|---|---|---|---|---|---|---|---|
| Risser and Ware (1999) | EEG | Simulated highway drive | Sleep apnea | Obstructive sleep apnea vs control group [between subject] | 30 | Lane position; Speed; Crash | The apnea group showed attention lapses more frequently across driving time compared to controls | Sleep apnea patients demonstrated increased lane position variability, crashes, and attention lapses. |
| Yang *et al.* (2010) | EEG | Play simulated automobile racing game | Epilepsy | N.A. | 91 | Steering wheel position; Velocity; Vehicle position; Crash frequency | Drivers' seizures detected and localised using EEG signal | Driving impairment during seizures differed in terms of both magnitude and character, depending on the seizure type |
| Krestel *et al.* (2011) | EEG | Avoid obstacle | Interictal epileptic activity (IEA) | N.A. | 25 | Reaction time | Individual epilepsy patients showed slower reaction times during IEA compared to EEG periods | Reaction time EEG could be used to assess fitness to drive; Generalised IEA of short duration seems to impair brain ability to react |
| Papageorgiou *et al.* (2012) | MRI | Drive while avoiding collision | homonymous visual field defects (unilateral vascular brain lesions) | Traffic density level & defect severity [between subject] | 26 | Collisions | Cortical structures associated with impaired collision avoidance identified: the parieto-occipital region and posterior cingulate gyrus in the right hemisphere and the inferior occipital cortex and parts of the fusiform (occipito-temporal) gyrus in the left hemisphere | No significant difference in collision avoidance between patients with left- and right-hemispheric lesions |
| Hung *et al.* (2014) | fMRI | Rural & urban driving | Focal cerebellar lesions | Healthy vs damaged brain [between subject] | 30 | Speed; Time to completion; Time to collision; Gap distance; Driving errors; Lane position | Cerebellar function is responsible for motor-speed coordination in basic driving manoeuvres; Cerebellar function is responsible for temporal-motor integration in complex driving situations | Drivers with cerebellar damage showed significantly compromised speed control during basic driving conditions; Complex driving ability is preserved after cerebellar damage due to functional compensation |

## Appendix D. Summary of studies on driver behaviour in semi-automated vehicles

| Reference | Signal type | Subject tasks | Design factors [design type] | Subjects | Behavioural measures | Imaging findings | Behavioural findings |
|---|---|---|---|---|---|---|---|
| Tsunashima and Yanagisawa (2009) | fNIRS | Simulated car following with and without adaptive cruise control | Active and non-active cruise control [within subject] | 4 | N.A. | Developing a signal processing method for analysing fNIRS signal of car drivers; Measuring brain activities of subjects with different level of mental calculation; Frontal lobe was less active during the adaptive cruise control; fNIRS images constructed with the proposed method agree to fMRI images in different workload levels | Outer portions of the frontal lobe were active when the subject drove without cruise control, indicating reduced activity related to driving performance |
| Arakawa *et al.* (2019) | NIRS | Drive on auto-pilot mode and take control after induced system failures | Manual drive, autonomous drive, system failure [between subject] | 13 | Eye movement; Distribution of body pressure on the seat; Blood pressure | Remarkable decline of the average relative haemoglobin concentration during the manual-driving scenario after encountering a system failure indicating that the driver may think he/she may make a mistake during manual driving; Mind distraction occurred prior to resuming control after a system failure because their brain activity at this instance was relatively low | Drivers' cognitive demand during autonomous driving is lower than that during manual driving; Drivers who depend on autonomous control systems experience stress upon switching to manual control after a system failure. |
| Cao *et al.* (2019) | EEG | Monotonous Cruise on a 4-lane highway and correct induced lane departures | Deviation onset, response onset, response offset [N.A.] | 27 | Reaction time | Providing dataset to develop methods for the design of individualised real-time neuro-ergonomic systems that enhance the situational awareness and decision-making of drivers, thereby improving human-system interactions | N.A. |
| Lee and Yang (2020) | EEG | Take over the control from the automated car | Take-over transition alert type [between subject] | 41 | N.A. | The visual-auditory-haptic warning scored the highest based on various EEG indexes, and was shown to be the most effective type of take-over transition alert | N.A. |

## Appendix E. Summary of studies on fatigued/drowsy drivers

| Reference | Signal type | Subject tasks | Subjects | Fatigue/performance indicators | Fatigue detection method | Observations | Validation findings/methods |
|---|---|---|---|---|---|---|---|
| Kar *et al.* (2010) | EEG | Simulated and field drive under different levels of fatigue | 40 | Subjective self-reported fatigue | Entropy measures in the wavelet domain | Five types of entropies (Shannon's entropy, Rényi entropy of order 2 and 3, Tsallis wavelet entropy and Generalized Escort-Tsallis entropy) could be used as indicators of driver fatigue | Subjective assessment of fatigue using standard questionnaire under field driving; Parameters vary in the same manner irrespective of simulated or actual driving; For the quantum of the signal variation, the effect of individual/contextual differences was significant |
| Zhao *et al.* (2011) | EEG | Monotonous simulated drive | 10 | Self-reported fatigue; Yawning; Rubbing; Driving performance; EOG | EEG-based algorithm to classify driver mental fatigue | Frontal, central and occipital signals extracted by multivariate autoregressive model can predict driver fatigue | KPCA–SVM algorithm enhances the generalisation ability of the classifier and improves the accuracy of driver mental fatigue recognition |
| Zhao *et al.* (2012) | EEG | Sustained simulated drive | 13 | EEG; ECG; Heart rate | EEG alpha and beta waves | Identifying possible indicators of driving fatigue: EEG alpha and beta, the relative power, P300 amplitude | The decline in the attention and arousal level as the driver gets fatigued |
| Chuang *et al.* (2015) | EEG | Monotonous cruise and correct induced lane departures | 10 | Reaction time | EEG-based perceptual function integration network (from multiple independent sources) | The parietal source classifier produced the highest accuracy among the five components of interest | The proposed integration network model outperformed conventional signal-based classifier |
| Huang *et al.* (2015) | EEG | Monotonous cruise and correct induced lane departures | 12 | Reaction time | Transfer entropy | Changes in effective connectivity in the cortico-cortical pathway is a neuropsychological signal for changes in alertness level | N.A. |
| Perrier *et al.* (2016) | EEG | Monotonous drive on real highway in sleepy and non-sleepy state | 24 | Standard deviation of lateral position (SDLP) | Fast Fourier Trans-forms (FFT) analysis of spectral power in three major EEG bands: alpha, beta, theta | SDLP and EEG signal (alpha and theta power spectra) increase dafter sleep deprivation and varied with time on task; Changes in SDLP and EEG did not correlate significantly; The effect of time on task on theta activity was more evident in sleep deprived driving | EEG is not appropriate to predict on-the-road driving performance. |
| Hu (2017a) | EEG | Monotonous simulated drive on a low-density highway | 28 | N.A. | Ensemble classifier: Fuzzy Entropy, Sample Entropy, Approximate Entropy, Spectral Entropy | N.A. | Fuzzy entropy and combined feature sets outperformed other feature sets |
| Hu (2017b) | EEG | Sustained attention simulated driving task on a highway with low density traffic | 12 | Subjective self-reported fatigue | Sample Entropy, Fuzzy Entropy, Approximate Entropy, Spectral Entropy | N.A. | Optimal performance of single channel is achieved using a combination of channel CP4, feature Fuzzy Entropy, and classifier Random Forest (RF). |
| Min *et al.* (2017) | EEG | Sustained attention simulated driving task on a highway with low density traffic | 12 | Subjective self-reported fatigue | Sample Entropy, Fuzzy Entropy, Approximate Entropy, Spectral Entropy | N.A. | The leave-one-out cross-validation approach obtained an accuracy of 98.3%, a sensitivity of 98.3% and a specificity of 98.2%. |
| Morales *et al.* (2017) | EEG | 2-hours simulated drive | 15 | Saccadic velocity;<br>Subjective (self-reported) alertness;<br>Speeding | Single-channel EEG | Power spectra of the delta EEG band showed an inverted U-shaped quadratic trend;<br>Power spectra of the beta band showed an increasing linear trend. | Driver's EEG power spectra and saccadic velocity changed over a 2-h simulated drive suggesting that the reduced level of arousal can be detected by the EGG signal |
| Nguyen *et al.* (2017) | EEG & NIRS | Awake and drowsy simulated drive | 11 | Blinking rate;<br>Eye closure;<br>Heart rate | Combination of EEG and NIRS | The oxyhemoglobin concentration change and the beta band power in the frontal lobe were found to be the most relevant indicators of the transition from awake to drowsy state | N.A. |
| Chuang *et al.* (2018) | EEG | Monotonous cruise and correct induced lane departures | 16 | Reaction time | Brain electrodynamics and hemodynamics | Observing strengthened alpha suppression in the occipital cortex, a common brain region of fatigue | Subjects were able to promptly respond to lane-deviation events, even if the sign of fatigue arose in the brain |
| Fonseca *et al.* (2018) | EEG | Monotonous cruise and correct induced lane departures | 17 | Reaction time | Transfer entropy | The spectral changes observed in the alertness oscillations can be explained by effective connectivity measures | Combining EEG, behavioural and actigraphy data can reveal new features of the decline in alertness |
| He *et al.* (2018) | EEG | Real expressway drive | 10 | N.A. | Power spectrum features of 14-channel EEG signal | Creating a brain network model;<br>Identifying a threshold to determine whether the brain network nodes are connected;<br>As fatigue increased, theta wave increased, and beta declined | As fatigue occurred, brain network continued to densify and neural activity increasingly synchronised across brain regions |
| Hu and Min (2018) | EEG | Monotonous simulated drive on a low-density highway | 22 | N.A. | Ensemble classifier: Fuzzy Entropy, Sample Entropy, Approximate Entropy, Spectral Entropy | N.A. | It is possible to use only one EEG channel to detect a driver fatigue state; The average highest recognition rate in this work was up to 94.0%, |
| Ahlström *et al.* (2019) | EEG | Simulated drive in rural and urban roads under alert and sleep deprived condition | 30 | Subjective (self-reported) fatigue; Line crossings | Lambda response | Sleep deprivation and time on task cause a general decrement in cortical responsiveness to incoming visual stimuli | Low lambda responses are associated with high subjective sleepiness and more line crossings, and could be used as a driver fatigue indicator |
| Barua *et al.* (2019) | EEG & EOG | Simulated rural (daylight and darkness) and suburban drive under alert and sleep deprived conditions | 30 | Subjective (self-reported) fatigue | EEG signal analysed by four different machine learning algorithms | Removing the ‘somewhat sleepy’ group and treating the classification as binary improves the result | The support vector machine showed better performance than the other classifiers 10% increase in accuracy when data from the individual being evaluated was included in the training dataset; Adding contextual information as features showed improvement in accuracy by 4% and 5% |
| Chen *et al.* (2019) | EEG | 1 hour field drive | 14 | Subjective (self-reported) fatigue | EEG signal decomposed to delta, theta, alpha and beta range by wavelet packet transform | Functional connectivity of the brain area was significantly different between the alert and fatigue states, particularly in alpha and beta range | The Support Vector Machine (SVM) achieved higher accuracy and outperformed the state-of-the-art systems |
| Gao *et al.* (2019) | EEG | Simulated drive under alert and fatigue state | 10 | Subjective (self-reported) fatigue | A recurrence network-based convolutional neural network method | N.A. | The proposed method can achieve an average accuracy of 92.95% and outperforms existing methods |
| Luo *et al.* (2019) | EEG | Simulated drive | 16 | Subjective (self-reported) fatigue | Multi-scale entropy feature extraction based on forehead EEG signal | Gender differences can affect fatigue detection | The accuracy of fatigue driving detection based on the forehead reached 95.37%; Adaptive multi-scale entropy has higher accuracy than single-scale entropy |
| Ma *et al.* (2019) | EEG | Simulated drive under alert and sleep deprived condition | 6 | Eye closure; Head nodding; Lane deviation | Integrated Principal Component Analysis & Machine Learning models | Parietal and occipital lobes were strongly associated with driver fatigue | Robust performance of the proposed modified method with classification accuracy up to 95%, which outperformed the conventional feature extraction |

## Appendix F. Summary of studies on driver brain activity related to decision-making tasks

| Reference | Signal type | Subject tasks | Phenomenon | Design factors [design type] | Subjects | Imaging findings | Behavioural findings |
|---|---|---|---|---|---|---|---|
| Callan *et al.* (2009) | fMRI | Turn right in left-hand traffic at a signalized intersection | Resolve uncertainty in decision making | Occluded & non-occluded view [within subject] | 14 | Resolving uncertainty reduced activity in the amygdala and anterior cingulate | Cost-weighted decision in driving is more pronounced than reward-weighted |
| Chein *et al.* (2011) | fMRI | Decide whether to stop (incur delay) or run the traffic light (risk crashing) | Risk taking | Adolescent, young adult & adult driver + Peer presence [between subject] | 40 | During peer observation blocks, adolescents selectively demonstrated greater activation in reward-related brain regions, indicative of increased risk taking; Areas of cognitive control were less recruited by adolescent drivers than adults | The presence of peers increases risk taking among adolescents but not adults |
| Haufe *et al.* (2011) | EEG | Drive a virtual racing car and avoid crash by emergency braking | Braking intention | N.A. | 18 | EEG potentials predict upcoming emergency braking; Prediction based on EEG signal outperforming the braking intention prediction based on pedal dynamics | N.A. |
| Haufe *et al.* (2014) | EEG | Real drive on a non-public test track | Braking intention | Drive only & drive with auditory task [within subject] | 20 | Demonstrating the feasibility of EEG-based emergency breaking intention detection | Results agreeing with those of the driving simulator study of Haufe *et al.* (2011) |
| Kim *et al.* (2014) | EEG | Simulated drive and respond to braking scenarios | Braking intention | Braking stimulus (e.g. soft vs sharp braking) [within subject] | 15 | Braking intention could be predicted via the EEG signal; Various types of braking intentions could be distinguished | Emergency braking is characterised by specific neural patterns of sensory perception and processing, motor preparation and execution |
| Vorobyev *et al.* (2015) | fMRI | Decide whether to stop (incur delay) or run the traffic light (risk crashing) | Risk taking | Personality trait & peer competition [between subject] | 34 | Risk-taking activated two areas in the left medial prefrontal cortex significantly more in low than in high risk-takers; Peer competition increased outcome-related activation in the right caudate head and cerebellar vermis in the entire sample | Decision to take risk activated adolescent brain more than the decision to stay safe; Social pressure (peer competition) was associated with longer decision time; Reward processing of risk-taking elevated under peer influence |
| Zhang *et al.* (2015) | EEG | Simulated & real drive and respond to direction | Turning intention | N.A. | 30 | Feasibility of decoding EEG signals to help estimate driver turning intention | N.A. |
| | | signs | | | | | |
| Foy *et al.* (2016) | fNIRS | Simulated drive | Car following & overtaking | Task type, traffic density, age, experience & gender [between subject] | 32 | Younger drivers had reduced prefrontal cortex activity compared to older drivers; Prefrontal cortex activity is associated with the mental workload of overtaking | No difference in the number of overtakes completed by younger and older drivers; Males overtook significantly more often than females |
| Hernández *et al.* (2018) | EEG | Simulated drive and respond to unexpected braking scenarios | Braking intention | Stress, workload and fatigue [within subject] | 7 | Successful recognition of the braking intention using the EEG signals classified based on SVM and CNN methods | Different braking reaction time detected across different experimental conditions |
| Tanida *et al.* (2018) | fMRI | Actively retrieve experiences in the past from episodic memory | Risk perception related to anticipatory control | Safe, risky and exciting drive episodes [within subject] | 14 | Brain areas corresponding to perceived safety indicate an anticipatory driving mode; An overlap of brain activation during mental imagery of safe compared to either risky or exciting car driving | Higher levels of arousal for the risky and the exciting drive when compared to the safe driving situation |
| Vecchiato *et al.* (2019) | EEG | Drive on a simulated traffic-free coastal road | Braking & acceleration intention | N.A. | 30 | Time-frequency analysis revealed a scalp pattern discriminating braking from acceleration; Theta signal can identify intention to brake 800 ms before onset of the event; The dorso-mesial premotor cortex is involved in the preparation of braking and acceleration | N.A. |

## Appendix G. Summary of studies on brain activity during general driving tasks

| Reference | Signal type | Investigated driving tasks | Subjects | Imaging findings | Behavioural implications |
|---|---|---|---|---|---|
| Walter *et al.* (2001) | fMRI | Active and passive simulated drive | 12 | Activity specifically associated with driving was found only in the sensorimotor cortex and the cerebellum;<br>Compared to passive driving, activity during driving was reduced in numerous brain regions | Simulated driving requires mainly perceptual-motor integration;<br>Limited cognitive capacity model of driving has to be revised |
| Calhoun *et al.* (2002) | fMRI | Active and passive simulated drive | 12 | Signal in the anterior cingulate cortex, an area often associated with error monitoring and inhibition, decreases exponentially with a rate proportional to driving speed;<br>Decreases in frontoparietal regions, implicated in vigilance, correlate with speed;<br>The neural correlates of driving are modulated by driving speed | A change in vigilance is initiated as the driving conditions begins;<br>During the fast drive condition, the vigilance component changes more; and the error correction and disinhibition component decreases faster. |
| Uchiyama *et al.* (2003) | fMRI | Active and passive car following (maintain constant distance) | 21 | Activation of the cerebellum may reflect visual feedback during smooth tracking of the preceding car;<br>Co-activation of the basal ganglia, thalamus and premotor cortex is related to movement selection;<br>Activation of a premotor-parietal network is related to visuo-motor coordination;<br>Anterior cingulate activity was related to error detection and response selection | N.A. |
| Horikawa *et al.* (2005) | PET | Active and passive simulated drive | 15 | Compared with the resting condition, simulated driving increased regional cerebral blood flow in the cerebellum, occipital, and parietal cortices | Increasing demands of vigilance and attention observed in poor performance drivers;<br>Significant correlation between driver performance (e.g. no. of crashes, time to complete the task) and the magnitude of activation in certain brain regions |
| Spiers and Maguire (2007) | fMRI | Free-roaming simulated drive in city environment | 20 | Prepared actions such as starting, turning, reversing and stopping were associated with a common network comprised of premotor, parietal and cerebellar regions;<br>Unexpected hazardous events such as swerving and avoiding collisions were associated with activation of lateral occipital and parietal regions;<br>Planning future actions and monitoring fellow road users were associated with activity in superior parietal, lateral occipital cortices and the cerebellum | Richness and dynamic nature of thought processes and actions associated with driving was observed |
| Mader *et al.* | fMRI | Passive simulated drive on | 16 | Common activations in frontal, parietal, temporal, occipital lobes, the | A familiar, monotonous route seem to lead to a |
| (2009) | | familiar & unfamiliar routes | | thalamus, and cerebellum was observed;<br>Significant activation for the unfamiliar route in the middle temporal and occipital cortex and in the cerebellum was observed | reduction in attention and perception processes which might be associated with a danger for commuters, even in specially trained drivers |
| Kan *et al.* (2013) | fMRI | Simulated driving events: straight drive, left turn, right turn, stopping | 16 | Robust maps of brain activity were obtained;<br>With care, fMRI of simulated driving is a feasible undertaking | N.A. |
| Yoshino *et al.* (2013) | fNIRS | Actual day & night drive on traffic-free expressway, parked, acceleration/deceleration, U-turns, maintain constant speed | 12 | Significant increase of cerebral oxygen exchange observed in frontal eye field;<br>Significant activation was detected during acceleration (deceleration) in right (left) frontal eye field | N.A. |
| Sakihara *et al.* (2014) | MEG | Pass viewing and active simulated driving | 14 | Power increase in the theta band was detected in the superior frontal gyrus (SFG) during active riving;<br>Power decreases in the alpha, beta, and low gamma bands in the right inferior parietal lobe, left postcentral gyrus, middle temporal gyrus, and posterior cingulate gyrus | Power changes during active driving indicative of increased attention, visuospatial processing and sensorimotor activity and object recognition |
| Oka *et al.* (2015) | fNIRS | Simulated drive on left and right curves | 15 | Cerebral activity in the right premotor cortex, right frontal eye field and bilateral prefrontal cortex was greater in left curve driving | Driver brain activity to be different when driving on left and right curves;<br>Left curve driving requires more attentional demand than right curves |
| Garcia *et al.* (2017) | EEG | Simulated drive and maintain constant lane and speed under sparse and heavy traffic | 28 | The proactive state is characterised by the delta-beta band;<br>The reactive state is characterised by the alpha band | Two neuro-behavioural states of brain activity during driving were identified;<br>The proactive brain state actively plans the response to the sensory information;<br>The reactive brain state processes incoming information and reacts to the environmental statistics |
| Unni *et al.* (2017) | fNIRS | Simulated drive while performing a memory tassk | 19 | Increased brain activation in bilateral inferior frontal and bilateral temporo-occipital areas as the working memory load increased | The memory task load level had significant effect on driver ability to maintain correct speed and their reaction times and braking variance;<br>Working memory load can have an effect on safety relevant driving behaviours |
| Di Flumeri *et al.* (2018) | EEG | Real drive under different road and traffic conditions | 20 | EEG signal can be used to assess driver mental workload | EEG-based workload index confirmed the significant impact of both traffic and road types on driver behaviour |
| Foy and Chapman (2018) | fNIRS | Simulated drive under different mental workload (road type) | 26 | A relation between mental workload and oxygenated haemoglobin in prefrontal cortex was found | Mental workload fluctuated during driving |
| Protzak and Gramann (2018) | EEG | Simulated and real drive | 35 | Increased P300 amplitudes reflected processing of infrequent and incorrect auditory feedback events in both the laboratory setting and the real-world setup | Demonstrating the possibility of investigating cognitive functions in highly artefactual driving scenarios |
| Yang *et al.* (2018a) | EEG | Simulated ordinary drive | 25 | Ordinary driving behaviour relates to all four brain regions, especially the temporal, occipital, and frontal regions | Acceleration, speed, and space headway may have potential correlation with EEG features |
| Yang *et al.* (2018b) | EEG | Simulated car following | 52 | A significant correlation between EEG patterns and car-following behaviour | Driving behaviour can be classified using EEG signal;<br>Five groups of driving behaviour were classified |
| Ding *et al.* (2019) | EEG | Simulated drive | 23 | N.A. | Based on EEG signal, the cognitive driving states were divided into four types: negative, calm, alert, and tense |
| Yan *et al.* (2019a) | EEG | Simulated drive on a circular curvy road | 23 | Driving style can be objectively classified using EEG signal | Different driving styles were related to different driving strategies and mental states |
| Yan *et al.* (2019b) | EEG | Simulated drive and turning tasks | 36 | The bilateral frontal gyrus was found to be activated when turning left and right | Correlation between driving behaviour, personality and EEG can be taken as a reference for the prediction of dangerous driving behaviour |
| Yang *et al.* (2019) | EEG | Simulated car following | 57 | The somatomotor region was found to have better predictive ability than other single brain source from the six brain regions, thus it may be more relevant to the driving states;<br>Full brain region features have better prediction performance than any single brain region features | EEG-based model has better performance than driving-data-based model;<br>Combination of driving features and EEG features, the hybrid approach, outperforms single-criterion approaches in short-term driving state prediction |